\documentclass[nohyper,12pt,letterpaper]{JHEP3}
\usepackage{epsfig}
\usepackage[latin1]{inputenc}
\usepackage{bbm,amsfonts}
\usepackage{graphicx}
\usepackage{amssymb,amsmath}

\author{Marco S. Bianchi$^\ast$,
  Matias Leoni$^{\dag}$,
  Andrea Mauri$^{\ddag \hash}$,
  Silvia Penati$^\dag$
  and Alberto Santambrogio$^\hash$\\\\
  $^\ast$Departamento de F\'isica, Universidad de Oviedo, Avda. Calvo Sotelo, 18, 33007 Oviedo, Espa\~na \\\\
  $^\dag$Dipartimento di Fisica, Universit\`a degli studi di Milano--Bicocca and
  INFN, Sezione di Milano--Bicocca, Piazza della Scienza 3, I-20126 Milano, Italy \\\\
  $^\ddag$Dipartimento di Fisica dell'Universit\`a degli studi di Milano, via Celoria 16, I-20133 Milano, Italy\\\\
  $^\hash$ INFN, Sezione di Milano, via Celoria 16, I-20133 Milano, Italy
  \qquad\\\\
  E-mail: \email{marco.bianchi@mib.infn.it, matias.leoni@mib.infn.it,
    andrea.mauri@mi.infn.it, silvia.penati@mib.infn.it,
    alberto.santambrogio@mi.infn.it }}

\abstract{For three dimensional $\mathcal{N}=6$ superconformal field theories we compute one--loop scattering amplitudes for any number of external particles. We focus on a particular subsector of ${\cal N}=2$ invariant superamplitudes for which the ordinary perturbative evaluation becomes very easy. The result we obtain is in general non--vanishing. For six external particles our findings are sufficient for determining the complete expression of the ${\cal N}=6$ superamplitude at this order. We discuss the symmetries of the result and its anomalous variation under superconformal generators.}

\preprint{April 2012\\ IFUM-995-FT\\FPAUO--12/07}

\title{
ONE LOOP AMPLITUDES IN ABJM
}


\csname @addtoreset\endcsname{equation}{section}

\def\bseq{\begin{subequation}}  
\def\eseq{\end{subequation}}
\def\bsea{\begin{subeqnarray}}  
\def\esea{\end{subeqnarray}}

\newcommand{\beq}{\begin{equation}}
\newcommand{\bea}{\begin{eqnarray}}
\newcommand{\eea}{\end{eqnarray}}
\newcommand{\eeq}{\end{equation}}

\newcommand {\non}{\nonumber}

\renewcommand{\a}{\alpha}
\renewcommand{\b}{\beta}

\renewcommand{\d}{\delta}
\newcommand{\pa}{\partial}
\newcommand{\g}{\gamma}
\newcommand{\G}{\Gamma}

\newcommand{\e}{\epsilon}

\renewcommand{\l}{\lambda}
\renewcommand{\L}{\Lambda}
\newcommand{\m}{\mu}

\newcommand{\n}{\nu}

\newcommand{\p}{\pi}

\newcommand{\s}{\sigma}

\newcommand{\Db}{\overline{D}}

\newcommand{\thb}{\overline{\theta}}

\renewcommand{\thb}{\overline{\theta}}

\def\Mb{\kern 2pt\mathchoice
        {
         \vbox{\hrule width10pt height 0.4pt depth 0pt
         \kern 1.2pt\hbox{\kern -2pt$\displaystyle M$}}}
        {
         \vbox{\hrule width10pt height 0.4pt depth 0pt
         \kern 1.2pt\hbox{\kern -2pt$\textstyle M$}}}
        {
\vbox{\hrule width6pt height 0.4pt depth 0pt
         \kern 1.0pt\hbox{\kern -2pt$\scriptstyle M$}}}
        {
         \vbox{\hrule width5pt height 0.4pt depth 0pt
         \kern 0.8pt\hbox{\kern -2pt$\scriptscriptstyle M$}}}}

\def\Sb{\kern 2pt\mathchoice
        {
         \vbox{\hrule width6pt height 0.4pt depth 0pt
         \kern 1.2pt\hbox{\kern -2pt$\displaystyle S$}}}
        {
         \vbox{\hrule width6pt height 0.4pt depth 0pt
         \kern 1.2pt\hbox{\kern -2pt$\textstyle S$}}}
        {
         \vbox{\hrule width3.5pt height 0.4pt depth 0pt
         \kern 1.0pt\hbox{\kern -2pt$\scriptstyle S$}}}
        {
         \vbox{\hrule width3pt height 0.4pt depth 0pt
         \kern 0.8pt\hbox{\kern -2pt$\scriptscriptstyle S$}}}}

\def\Rb{\kern 2pt\mathchoice
        {
         \vbox{\hrule width5.5pt height 0.4pt depth 0pt
         \kern 1.2pt\hbox{\kern -2.5pt$\displaystyle R$}}}
        {
         \vbox{\hrule width5.5pt height 0.4pt depth 0pt
         \kern 1.2pt\hbox{\kern -2.5pt$\textstyle R$}}}
        {
         \vbox{\hrule width3.5pt height 0.4pt depth 0pt
         \kern 1.0pt\hbox{\kern -2.2pt$\scriptstyle R$}}}
        {
         \vbox{\hrule width3pt height 0.4pt depth 0pt
         \kern 0.8pt\hbox{\kern -2.2pt$\scriptscriptstyle R$}}}}

  \def\pp{{\mathchoice
      {
          \kern 1pt%
          \raise 1pt
          \vbox{\hrule width5pt height0.4pt depth0pt
            \kern -2pt
            \hbox{\kern 2.3pt
              \vrule width0.4pt height6pt depth0pt
              }
            \kern -2pt
            \hrule width5pt height0.4pt depth0pt}%
            \kern 1pt
       }
        {
          \kern 1pt%
          \raise 1pt
          \vbox{\hrule width4.3pt height0.4pt depth0pt
            \kern -1.8pt
            \hbox{\kern 1.95pt
              \vrule width0.4pt height5.4pt depth0pt
              }
            \kern -1.8pt
            \hrule width4.3pt height0.4pt depth0pt}%
            \kern 1pt
        }
        {
          \kern 0.5pt%
          \raise 1pt
          \vbox{\hrule width4.0pt height0.3pt depth0pt
            \kern -1.9pt  
            \hbox{\kern 1.85pt
              \vrule width0.3pt height5.7pt depth0pt
              }
            \kern -1.9pt
            \hrule width4.0pt height0.3pt depth0pt}%
            \kern 0.5pt
        }
        {
          \kern 0.5pt%
          \raise 1pt
          \vbox{\hrule width3.6pt height0.3pt depth0pt
            \kern -1.5pt
            \hbox{\kern 1.65pt
              \vrule width0.3pt height4.5pt depth0pt
              }
            \kern -1.5pt
            \hrule width3.6pt height0.3pt depth0pt}%
            \kern 0.5pt
        }
    }}

  \def\mm{{\mathchoice
               {
                 \kern 1pt
           \raise 1pt    \vbox{\hrule width5pt height0.4pt depth0pt
                  \kern 2pt
                  \hrule width5pt height0.4pt depth0pt}
                 \kern 1pt}
               {
                \kern 1pt
           \raise 1pt \vbox{\hrule width4.3pt height0.4pt depth0pt
                  \kern 1.8pt
                  \hrule width4.3pt height0.4pt depth0pt}
                 \kern 1pt}
               {
                \kern 0.5pt
           \raise 1pt
                \vbox{\hrule width4.0pt height0.3pt depth0pt
                  \kern 1.9pt
                  \hrule width4.0pt height0.3pt depth0pt}
                \kern 1pt}
               {
               \kern 0.5pt
         \raise 1pt  \vbox{\hrule width3.6pt height0.3pt depth0pt
                  \kern 1.5pt
                  \hrule width3.6pt height0.3pt depth0pt}
               \kern 0.5pt}
               }}

\def\pd{{\kern0.5pt
           + \kern-5.05pt \raise5.8pt\hbox{$\textstyle.$}\kern
0.5pt}}

\def\pmd{{\kern0.5pt
          \pm \kern-5.05pt
\raise6.3pt\hbox{$\textstyle.$}\kern1.5pt}}

\def\md{{\mathchoice
   {
      {{\kern 1pt - \kern-6.2pt \raise5pt\hbox{$\textstyle.$}\kern
1pt}}}
    {
      {{\kern 1pt - \kern-6.2pt \raise5pt\hbox{$\textstyle.$}\kern
1pt}}}
    {
      {\kern0.5pt - \kern-5.05pt
\raise3.4pt\hbox{$\textstyle.$}\kern0.5pt}}
    {
      {\kern0.5pt - \kern-5.05pt
\raise3.4pt\hbox{$\textstyle.$}\kern0.5pt}}}}

\def\beq{\begin{equation}}
\def\eeq{\end{equation}}
\def\bea{\begin{eqnarray}}
\def\eea{\end{eqnarray}}
\def\Tr{\mathrm{Tr}}

\def\a{\alpha}
\def\b{\beta}
\def\g{\gamma}

\def\d{\delta}
\def\e{\epsilon}

\def\th{\theta}
\def\l{\lambda}
\def\G{\Gamma}

\def\L{\Lambda}

\begin{document}

\section{Introduction}

The three dimensional version of the AdS/CFT correspondence \cite{ABJM, ABJ} which states the equivalence between a ${\cal N}=6$ superconformal, $U(N)_K \times U(N)_{-K}$  quiver Chern--Simons--matter theory and a type IIA string theory on ${\rm AdS}_4 \times CP^3$ or M--theory on ${\rm AdS}_4 \times S^7/Z_K$, provides an alternative, non--trivial arena where studying the deep nature of the correspondence.
In fact, the theories appearing on the two sides of the correspondence exhibit quite different features compared to their four dimensional counterparts, so they might disclose novel aspects.

On both sides of the correspondence, integrable structures seem to emerge in the planar limit. In fact,  at strong coupling  the classical integrability of the string non--linear sigma model has been argued
\cite{Arutyunov:2008if,Stefanski:2008ik,Sorokin:2010wn} and tree level string scattering amplitudes have been proven to factorize \cite{Kalousios:2009ey}. At weak coupling, the dilatation operator for gauge invariant, local, composite operators has been related to the Hamiltonian of an integrable spin chain \cite{MZ} and an all--loop Bethe ansatz for determining the spectrum of the anomalous dimensions has been proposed \cite{GV} which is consistent with the $Osp(6,4)$ algebraic curve at strong coupling \cite{GV2}, agrees with the exact S--matrix conjectured in \cite{AN} and matches the spectrum of type IIA string theory on ${\rm AdS}_4 \times CP^3$ in the Penrose limit \cite{Astolfi:2011bg}.
Moreover,  the dispersion relation for magnons has been computed in terms of a non--trivial function of the 't Hooft parameter that interpolates between strong and weak coupling results \cite{Nishioka:2008gz}--\cite{LMMSSST}.

However, a complete comprehension of integrability at quantum level has not been reached yet and further investigation is required.

On the field theory side, integrable structures are expected to have important  consequences on its on--shell sector. In particular, the existence of an infinite algebra of non--local conserved currents, the Yangian \cite{Drummond:2009fd}, would constrain the form of the scattering amplitudes and their dualities with other important quantities like Wilson loops \cite{Drummond:2007aua}--\cite{Drummond:2007cf} and correlation functions \cite{AEKMS,EKS}. Therefore, a direct study of the properties of scattering amplitudes can be used for grasping further indications of the integrable structure underlying the planar sector of the theory.

At tree level, quite a number of well--established results are now available. General constraints coming from requiring superconformal invariance, once solved, allow to determine tree level superamplitudes in terms of a restricted number of independent functions \cite{BLM}. Explicit results have been found for the four and six--point amplitudes and their invariance under level one Yangian generators has been proven \cite{BLM,HL2, HL}.  Dual superconformal invariance \cite{Drummond:2008vq} of all tree--level amplitudes has been subsequently proven \cite{GHKLL} by exploiting a three dimensional version of the the BCFW recursion relations \cite{BCFW}.  Finally, a generating function for scattering amplitudes has been proposed in \cite{Lee}  that is manifestly Yangian invariant.

Quantum investigation of these properties passes necessarily through the difficult task of computing perturbative corrections to the scattering superamplitudes. At loop level, very little has been done so far.
Explicit results are available only for four--point amplitudes. The complete superamplitude is one--loop vanishing \cite{ABM}--\cite{BLMPS1}, while an interesting non--trivial contribution has been found at two loops in the planar limit \cite{CH, BLMPS1, BLMPS2} that has a number of remarkable properties. When divided by its tree--level counterpart, it is dual superconformal invariant and coincides with the second order expansion of a light--like four--polygon Wilson loop \cite{HPW}. This gives indication that a Wilson loop/scattering amplitude duality might be at work even if this duality  does not have a clear proof at strong coupling yet \footnote{Attempts to mimic what happens in four dimensions \cite{BM} have experimented the appearance of singularities in the fermionic T--transformations \cite{ADO}--\cite{Colgain:2012ca}.}.
The two--loop result can be thought of as the lowest order expansion of an exponentiation formula for the all--loop amplitude which can be justified via AdS/CFT correspondence \cite{BLMPS2} by adapting to the case of type IIA string in ${\rm AdS}_4 \times CP^3$ the Alday--Maldacena prescription \cite{AM} for computing scattering amplitudes at strong coupling. Finally, up to scheme dependent and subleading terms in the IR regulator, it has been proven to be equal to the four--point amplitude of ${\cal N}=4$ SYM theory at one--loop \cite{BLP}, so giving further support to the correctness of the exponentiation proposal.

Beyond four--point amplitude, nothing is known at quantum level. The scope of this paper is to provide a first non--trivial result for higher points planar amplitudes at one loop.

As is well known, scattering amplitudes involve only matter particles and their number is constrained by gauge invariance to be even. Introducing an on--shell superspace formalism, it is possible to construct superamplitudes and classify them in terms of their grassmannian degree. It follows that for $n$ external particles the degree is that of an ${\rm N}^{(n/2 - 2)}$MHV superamplitude. At four points they correspond to the MHV case while, starting from $n=6$, we are not dealing with MHV amplitudes anymore, so the kind of expected properties will be different from the ones of the four--point amplitudes emphasized above. In particular, no duality is expected with bosonic light--like polygon Wilson loops that are one--loop vanishing in three dimensional Chern--Simons (matter) theories \cite{HPW, BLMPRS}.

Working in ${\cal N}=2$ superspace, in the large $N$ limit, we concentrate on particular subsectors of $n$--point amplitudes for which an ordinary perturbative approach based on  Feynman super--diagrams is feasible at one loop, given the small number of contributions allowed at this order. 
We derive iterative formulae for both the tree--level and one--loop contributions which are valid for any number of external superfields. The  various component amplitudes may be straightforwardly extracted from them. We provide some all--$n$ formulae for the simplest components, that is the ones involving the greatest number of scalars or fermions.

While for generic $n$ our results do not cover all kinds of amplitudes one can construct, for the special $n=6$ case our findings allow for reconstructing the complete superamplitude at one loop.  

The results we find exhibit regions of discontinuity in momentum space. For instance, for $n=6$ it is proportional to the sum of two kinematic factors which take only $\pm 1$ values. Therefore, there are physically accessible regions of momentum space where the six--point amplitude vanishes and regions where it does not vanish. Different regions are separated by discontinuities that correspond to configurations where two adjacent momenta become collinear.

The result for the six--point amplitude is proportional to sign functions. When acting with a tree--level generator of superconformal transformations it gives rise to a delta--function, signaling the appearance of an anomaly at one loop which resembles the tree--level holomorphic anomaly in four dimensions.

Our calculation can be easily generalized to the ABJ theory \cite{ABJ} corresponding to a more general $U(M)_K \times U(N)_{-K}$ gauge group. We provide the explicit result for the six--point amplitude and discuss its discontinuities. In this case, the amplitude is always non--vanishing, although in some regions of momentum space it is proportional to the color factor $(M+N)$, whereas in other regions it is proportional to the parity violating factor $(M-N)$. Still, going from one region to another requires adjacent momenta to become collinear.

The paper is organized as follows. In Section 2 we summarize generalities on the scattering superamplitudes and corresponding component amplitudes in ABJ(M) theory, discussing the peculiar subsectors we restrict to. In Section 3 we present the detailed calculation for the ${\cal N}=2$, six--point superamplitude, whereas in Section 4 we give the general result for $n$--point superamplitudes. Finally, in Section 5 we discuss the relevance of the result and its properties.  Three Appendices follow that list our set of conventions and give technical details of the calculation.

\vskip 15pt
{\bf Note added:} During completion of this work we were informed about another paper, appearing in the arXiv the same day, which has partial overlapping with our results \cite{BBLM}. 
In that paper, using a different approach, the authors obtain the same result for the one--loop six--point amplitude.

\section{Generalities on scattering amplitudes in ABJ(M)}

The ABJ(M) theories \cite{ABJM, ABJ} possess ${\cal N}=6$ supersymmetry, as the corresponding actions exhibit $SO(6)_R$ symmetry when written in components with the auxiliary fields set on--shell \cite{Klebanov}. For $U(M)_K \times U(N)_{-K}$  group, they involve two gauge vector multiplets
each of them in the adjoint representation of one of the two gauge groups, four complex scalars and their fermionic partners  $(\phi^I, \psi_I)$, $I =1, \cdots , 4$ in the bifundamental $(M, \bar{N})$ representation and their conjugates $(\bar{\phi}_I, \bar{\psi}^I)$ in the $(\bar{M}, N)$ representation. The gauge sector is described by a two--level Chern--Simons action, so the gauge fields are not propagating and cannot enter scattering processes.

The only non--trivial amplitudes of the theory are those involving matter external particles. We classify as particles the ones carrying $(M, \bar{N})$ indices and antiparticles the ones carrying $(\bar{M}, N)$ indices.

Each external particle carries an on--shell momentum $p_{\a\b}$ ($p^2 =0$), polarization  spinor $\l_\a$ for fermions, an $SU(4)$ index and color indices corresponding to the two gauge groups.  The on--shell condition for the momentum can be explicitly solved by expressing $p_{\a\b} = \l_\a \l_\b$, in terms of SL(2, $\mathbb{R}$) commuting spinors.

Scattering superamplitudes can be written in an on--shell ${\cal N}=3$ superspace formalism \cite{BLM}. Breaking the $SU(4)$ R--symmetry group down to $U(3)$ and introducing a set of three Grassmann coordinates, $\eta^{A}$, $A=1,2,3$,  in the fundamental representation of $SU(3)$, the matter fields can be embedded into two multiplets, a scalar $\Phi$ and a fermion $\bar{\Phi}$, according to
\begin{align}\label{spectrum}
\Phi(\Lambda)
&= \phi^4(\lambda)
  +\eta^A\,\psi_A(\lambda)
  +\frac12\,\epsilon_{ABC}\,\eta^A\,\eta^B\,\phi^C(\lambda)
  +\frac{1}{3!}\,\eta^A\,\eta^B\,\eta^C\,\epsilon_{CBA}\,\psi_4(\lambda)\non\\
\bar{\Phi}(\Lambda)
&= \bar\psi^4(\lambda)
  +\eta^A\,\bar\phi_A(\lambda)
  +\frac12\,\epsilon_{ABC}\,\eta^A\,\eta^B\,\bar\psi^C(\lambda)
  +\frac{1}{3!}\,\eta^A\,\eta^B\,\eta^C\,\epsilon_{CBA}\,\bar\phi_4(\lambda)
\end{align}
where we have defined $\L \equiv (\l, \eta)$.
The former superfield contains the particles, whilst the second one the antiparticles.

In terms of these superfields a generic superamplitude has the form
\begin{equation}
{\cal A}_n\,(\Phi^{a_1}_{1 \; \bar{a}_1}, \, \bar{\Phi}^{\bar{b}_2}_{2 \; b_2},\, \Phi^{a_3}_{3 \; \bar{a}_3} ,\ldots,\bar{\Phi}^{\bar{a}_n}_{n \; a_n})
\end{equation}
From ${\cal A}_n$, the component amplitudes can be read as the coefficients of  its $\eta$--expansion.

This formulation does not allow for a direct evaluation of the superamplitudes, since no Lagrangian is known for the $\Phi, \bar{\Phi}$ superfields. However, in \cite{BLM} it has been shown that requiring
$Osp(6|4)$ superconformal invariance  of the superamplitudes restricts them to be of the form
\beq
\label{general}
{\cal A}_n(\L_1, \cdots , \L_n) = \d^{(3)} (P) \, \d^{(6)}(Q) \, \sum_{k=1}^K  f_{n,k} \, F_{n,k}
\eeq
where we recognize the delta functions for the supermomentum conservation. The non--trivial part is given in terms of R--symmetry invariants $F_{n,k}$ whose number equals the number of singlets in the representation $(4 \oplus \bar{4})^{\otimes (n-4)}$ (see Ref. \cite{BLM} for details).

The calculation of the superamplitudes is then reduced to the determination of the coefficients $f_{n,k}$ order by order in perturbation theory.
As discussed, at least in the simple cases of four and six--point superamplitudes, these coefficients can be inferred from the knowledge of a restricted number of component amplitudes. It is then sufficient to develop an efficient way for computing few independent component amplitudes.

To accomplish that, we find convenient to work in ${\cal N}=2$ superspace where the physical spectrum is organized in terms of two gauge multiplets and four matter chiral superfields $A^i$ and $B_i$, $i=1,2$ \cite{Klebanov}. Using the same capital letter to indicate their bosonic components and Greek letters $\a^i$ and $\b_i$ for the fermionic components, the dictionary for mapping ${\cal N}=3$ superfields (\ref{spectrum}) to ${\cal N}=2$ superfields is
\beq
\phi^A \rightarrow \left( A^1, A^2, \bar B^1, \bar B^2 \right)
\qquad\qquad\bar \phi_A \rightarrow \left( \bar A_1, \bar A_2, B_1, B_2 \right)
\eeq
and similarly for fermions,
\beq
\psi_A \rightarrow \left( -\a^B\,\e_{BA}\,e^{-i\p/4}\,\,;\,\, \bar\b^B\,\e_{BA}\,e^{i\p/4} \right)
\ \ \ \bar \psi^A \rightarrow \left(  -\e^{AB}\,\bar\a_B\,e^{i\p/4}\,\,;\,\, \e^{AB}\,\b_B\,e^{-i\p/4} \right)
\eeq
The action for the ABJ(M) theory written in ${\cal N}=2$ formalism can be found in Appendix B.

In terms of these new superfields the general expression for a color ordered ${\cal N}=2$ superamplitude is
\beq
{\cal A}_n \left( X^{a_1}_{1 \; \bar{a}_1}\,  \bar{X}^{\bar{b}_2}_{2 \; b_2} \, X^{a_3}_{3 \; \bar{a}_3} \cdots 
\bar{X}^{\bar{b}_n}_{n \; b_n} \right) = \left(\frac{4\pi}{K}\right)^{\frac{n}{2}-1}
  \sum_{\s}
\mathcal{A}_n(\s(1),   \cdots , \s(n) ) \; \d^{a_{\s(1)}}_{b_{\s(2)}} \, \d^{\bar{b}_{\s(2)}}_{\bar{a}_{\s(3)}}
\cdots \d^{\bar{b}_{\s(n)}}_{\bar{a}_{\s(1)}}
\eeq
where $X$ stays for any of the $(A^i, \bar{B}_i)$ superfields and $\bar{X}$ for any of their hermitian conjugates.
Here the sum is over exchanges of even and odd sites among themselves, up to cyclic permutations by two sites. In this way we can forget about the color factor and concentrate on a particular color ordered coefficient ${\cal A}_n$ that will be determined perturbatively as power series in the effective couplings $\l = M/K$ and $\hat \l = N/K$. We will always work in the large $M$, $N$ limit.

Loop contributions to the ${\cal A}_n$ superamplitude can be read from loop corrections to terms of the effective action proportional to $\int d^4 \th  \, {\rm Tr} ( X_1 \bar{X}_2 X_3 \cdots \bar{X}_n)$ with possible spinorial derivatives acting on the fields. Applying the $d^4 \th$ integration will give rise to non--trivial component amplitudes. 

We obtain the effective action performing D--algebra manipulations and momentum integrals in Euclidean metric. Successively, we  Wick rotate and analytically continue the final result to the mostly plus Minkowskian signature (see Appendix A for details).

Given the particular structure of the interaction vertices that can be read from the action (\ref{action}), the number of diagrams entering the evaluation of an amplitude depends drastically on the particular configuration of the external superfields.

We have managed to select a particular subsector of color ordered superamplitudes that at tree level are given by a single diagram featuring superpotential interactions only, and at one loop, in the planar limit, get corrections only from a single topology of diagrams, that is a box diagram. The key ingredient for selecting such superamplitudes is to avoid adjacent fields to share the same $SU(2)_A\times SU(2)_B$ flavor index.  One particularly simple choice is the following
\beq
\label{4n+2}
{\cal A}_{4m+2} ((A^2 \bar{A}_1)^{m} \bar{B}^1  (\bar{A}_2 A^1)^{m} B_1)
\eeq
\beq \label{4n}
{\cal A}_{4m} (A^1 (\bar{A}_2 A^1)^{(m-1)} B_1 A^2 (\bar{A}_1 A^2)^{(m-1)} B_2)
\eeq
or their cyclic permutations. For $m=1$  (\ref{4n}) reduces to the four--point chiral superamplitude computed in \cite{BLMPS1, BLMPS2}.

Since classically the theory is invariant under a $SU(2)_A \times SU(2)_B$ global symmetry and a $Z_2$ symmetry that exchanges $U(M) \leftrightarrow U(N)$, $K \leftrightarrow -K$, $V \leftrightarrow \hat{V}$ and $A^i \leftrightarrow B_i$,  the particular choice of flavors we have made in the previous expressions is not restricting. Applying a $SU(2)_A \times SU(2)_B$ transformation  we will obtain amplitudes  with $A^1$ and $A^2$ and/or $B_1$ and $B_2$ interchanged. Similarly, applying a $Z_2$ transformation we will obtain amplitudes with $A$ and $B$ interchanged.

\section{The complete six--point superamplitude at one loop}\label{sec:sixpoint}

We concentrate on the particular color ordered NMHV superamplitude that we obtain from (\ref{4n+2}) setting $m=1$, that is
${\cal A}_{6}( A^1 B_1 A^2 \bar{A}_1 \bar{B}^1 \bar{A}_2)$. We then look for perturbative contributions to terms in the effective action of the form
\beq
 \int d^4 \th  \; {\rm Tr} (A^1 B_1 A^2 \bar{A}_1 \bar{B}^1 \bar{A}_2)
\eeq
We perform the calculation in the general $U(M)_K \times U(N)_{-K}$ case, in the planar limit. As already mentioned, we work in Euclidean superspace and only at the end we will rotate back to Minkowski to obtain the physical amplitudes. Conventions for ${\cal N}=2$ Euclidean superspace are given in Appendix B, whereas the prescription for analytically continue the result to Minkowski can be found in Appendix A.
\FIGURE{ 
    \centering
    \includegraphics[width=1.\textwidth]{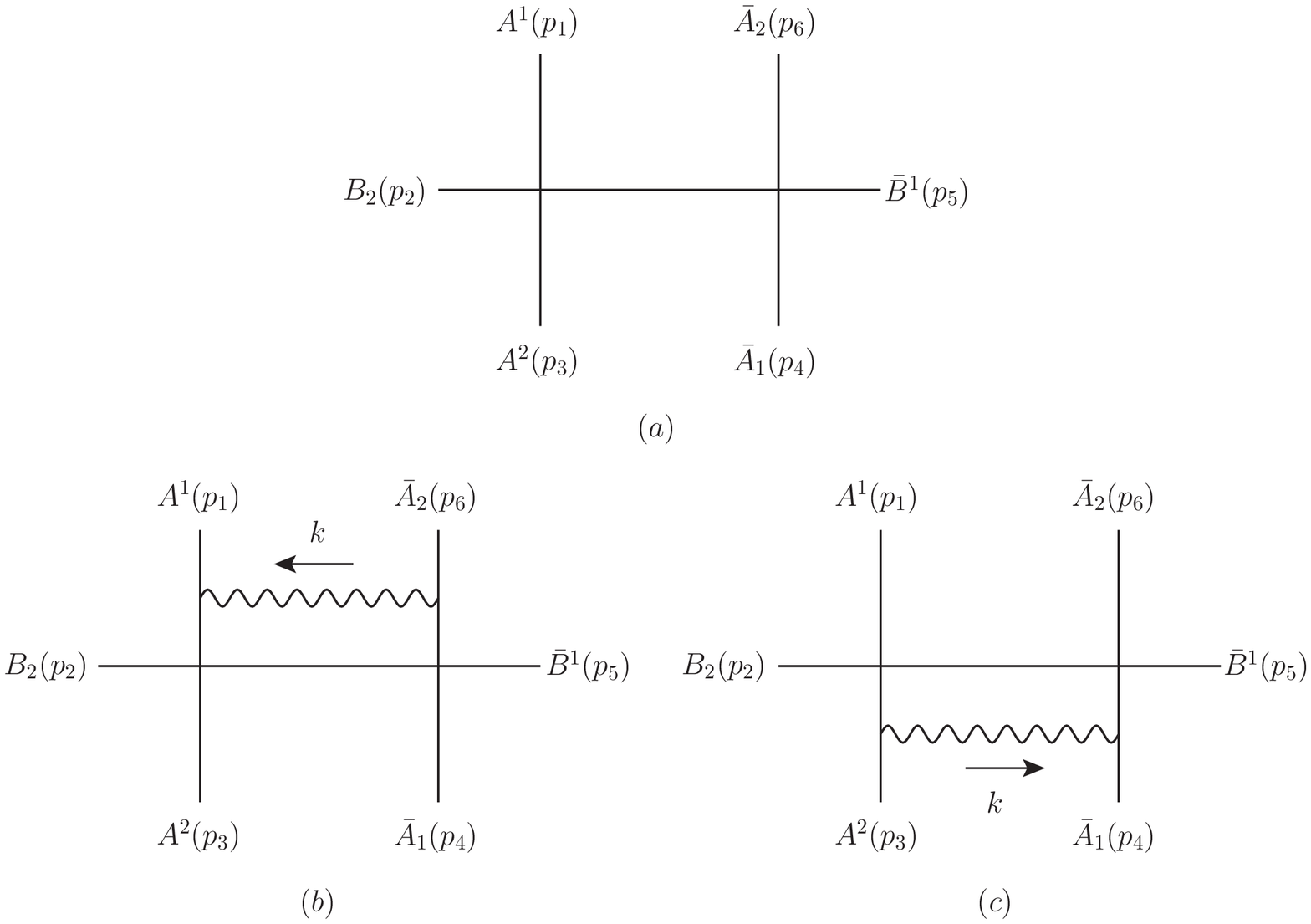}
    \caption{Diagrams contributing to the tree level and one--loop effective action with six external superfields. In diagram $(b)$ the wavy line corresponds to a $\langle V V\rangle$ propagator, whereas in diagram $(c)$ it corresponds to a $\langle \hat{V} \hat{V}\rangle$ one.}
    \label{fig:6points}
}

At tree level, the only contribution to this amplitude is drawn in Figure \ref{fig:6points}$(a)$. Assigning outgoing momenta $p_1, \cdots , p_6$ starting from the upper left leg and going counterclockwise,  it corresponds to the superspace integral
\begin{equation}
\Gamma_6^{tree} = -
\int d^4{\theta}
\frac{\mathrm{Tr}\left(A^1(p_1) B_1(p_2) A^2(p_3)\bar{A}_1(p_4)\bar{B}^1(p_5)\bar{A}_2(p_6)\right)}
{(p_1+p_2+p_3)^2}
\end{equation}
where we omit the integration over all the external momenta and the conservation delta function.

Taking into account the free equations of motion (\ref{onshell}) and the chirality conditions, and using the projections (\ref{components}), from this expression we can read the component amplitudes by integrating on the spinorial variables. In particular, we find a non--vanishing purely scalar amplitude, while the purely fermionic one is trivially zero
\begin{equation}
\label{tree}
\mathcal{A}_{6,s}^{(0)}(A^1 B_1 A^2\bar A_1\bar B^1\bar A_2) = 1
 \qquad ; \qquad \mathcal{A}_{6,f}^{(0)}(\alpha^1\,\beta_1\,\alpha^2\,\bar \alpha_1\,\bar \beta^1\,\bar \alpha_2)=0
\end{equation}
Written in terms of the components of $SU(4)$ multiplets these amplitudes are
\begin{equation}
\mathcal{A}_{6,s}^{(0)}(\phi^1\bar\phi_3\phi^2\bar\phi_1\phi^3\bar\phi_2) = 1
 \qquad ; \qquad \mathcal{A}_{6,f}^{(0)}(\psi_2\bar\psi^4\psi_1\bar\psi^2\psi_4\bar\psi^1)=0
\end{equation}

At one loop, planar contributions correspond to the two box diagrams in Figure \ref{fig:6points}$(b)$ and \ref{fig:6points}$(c)$. We perform D--algebra on these diagrams in order to reduce the spinorial derivatives inside the loop to a single $D^2 \bar{D}^2$ factor. This guarantees the entire expression to be local in the $\theta$--variables and  proportional to a  sum of ordinary momentum integrals. In both cases, we are left with one triangle and one box integral which eventually can be expressed in terms of triangles (see Appendix A).

Diagram \ref{fig:6points}$(b)$ where a $V$--vector propagates, gives
\begin{align}
\label{intermediate}
\Gamma^{1-loop}_{6(b)} &= -4 \pi\, \l 
\int d^4\theta\, \frac{d^3k}{(2\pi)^3}  \non\\&
\frac{k^2\, \mathrm{Tr}\,\left( D^{\a} A^1\, B_1\, A^2\, \bar{A}_1\, \bar{B}^1\,\bar{D}_{\a} \bar{A}_2) +  2\,\e_{\mu\nu\rho}\, k^{\mu}\, p_{1}^{\nu}\,p_{6}^{\rho}\, \mathrm{Tr}\,(A^1\, B_1\, A^2\,\bar{A}_1\,\bar{B}^1\,\bar{A}_2\right)}
{k^2\,(k-p_1)^2\,(k-p_1-p_2-p_3)^2\,(k+p_6)^2}
\end{align}
In Euclidean space both integrals are finite and given in (\ref{integral1}, \ref{integral2}).

Summing the two contributions and using integration by parts and on--shell conditions (\ref{onshell}) we can simplify the result to
\begin{equation}
\Gamma^{1-loop}_{6(b)} = \frac{\pi}{2} \, \l 
  \int d^4\theta \,\frac{\mathrm{Tr}\left(D^{\a} A^1\, D^{\b} B_1\, D_{\b} A^2\,\bar{D}^{\g}\bar{A}_1\,\bar{D}_{\g}\bar{B}^1\,\bar{D}_{\a}\bar{A}_2 \right)
}{p_{123}^2\sqrt{p_{23}^2}\sqrt{p_{45}^2}\sqrt{p_{16}^2}}
\end{equation}
where $p_{123}^2 \equiv (p_1 + p_2 + p_3)^2$. 

Performing a similar calculation, from diagram \ref{fig:6points}$(c)$ where a $\hat{V}$--vector propagates we obtain
\begin{equation}
\Gamma^{1-loop}_{6(c)} = -\frac{\pi}{2} \, \hat \l 
  \int d^4\theta \,\frac{\mathrm{Tr}\left(D^{\a} A^1\,D_{\a} B_1 \,D^{\b} A^2\,\bar{D}_{\b}\bar A_1\,\bar{D}^{\g}\bar B^1\,\bar{D}_{\g}\bar A_2
\right)}
{p_{123}^2\sqrt{p_{12}^2}\sqrt{p_{34}^2}\sqrt{p_{56}^2}}
\end{equation}
At this point we analytically continue the result to Minkowski space where we interpret $\sqrt{p^2} \equiv \sqrt{p^2 + i\e}$.

Projecting to components the total effective action given by the sum of contributions $1(b)$ and $1(c)$, we obtain scattering amplitudes for scalars and fermions. Once again, the simplest ones are those made out of only scalars and only fermions. For the scalar one, going back to ${\cal N}=3$ superspace notation, it is easy to realize that 
\begin{equation}
\label{scalar2}
\mathcal{A}_{6,s}^{(1)}(A^1 B_1 A^2\bar A_1\bar B^1\bar A_2) \, \equiv \, \mathcal{A}_{6,s}^{(1)}(\phi^1\bar\phi_3\phi^2\bar\phi_1\phi^3\bar\phi_2)=0
\end{equation}
whereas for the purely fermionic one, we obtain
\begin{align}
\label{fermion2}
&\mathcal{A}_{6,f}^{(1)}(\alpha^1\,\beta_1\,\alpha^2\,\bar \alpha_1\,\bar \beta^1\,\bar \alpha_2)
\, \equiv \, -\mathcal{A}_{6,f}^{(1)}(\psi_2\bar\psi^4\psi_1\bar\psi^2\psi_4\bar\psi^1)=\nonumber\\
& \frac{\pi}{2}\, 
\left[ \l\; 
\frac{\langle1\, 2\rangle}{\sqrt{\langle1\, 2\rangle^2}} \, \frac{\langle3\, 4\rangle}{\sqrt{\langle3\, 4\rangle^2}}
\, \frac{\langle5\, 6\rangle}{\sqrt{\langle5\, 6\rangle^2}}
+ \hat \l\; 
\frac{\langle2\, 3\rangle}{\sqrt{\langle2\, 3\rangle^2}} \, \frac{\langle4\, 5\rangle}{\sqrt{\langle4\, 5\rangle^2}}
\, \frac{\langle6\, 1\rangle}{\sqrt{\langle6\, 1\rangle^2}}  \right] \equiv {\cal C}(P)
\end{align}
We see the appearance of $\frac{\langle i \;  i+1 \rangle}{\sqrt{\langle i \;  i+ 1 \rangle^2}}$ ratios that give rise to sign functions. Therefore, as it will be discussed in detail in Section 5, the expression inside the brackets is constant and proportional to $(\l +\hat{\l})$ or $(\l-\hat{\l})$, according to the particular choice of the kinematic configuration.  In particular, for ABJM theory ($\l=\hat{\l}$) the amplitude  is either proportional to $2\l$ or exactly $0$. The discontinuities correspond to regions of collinearity for two adjacent momenta.

\vskip 20pt

The one--loop results (\ref{scalar2}, \ref{fermion2}) we have obtained are sufficient for reconstructing the complete ${\cal N}=6$ superamplitude, as we are going to discuss.

For $n=6$, eq. (\ref{general}) reduces to \cite{BLM}
\beq\label{superamp4}
{\cal A}_6 = \d^3(P)\, \d^6(Q)\, \left[ f^+(\l) \, \d^3(\a) + f^-(\l) \, \d^3(\b) \right]
\eeq
where $f^{\pm}(\l)$ are functions to be determined through an explicit computation and the two independent R--invariant functions
are given in terms of spinorial variables $\a^A \equiv x^+ \cdot \eta^A$, $\b^A \equiv x^- \cdot \eta^A$ with \cite{BLM}
\begin{align}
\label{pm}
 x^\pm_{i}&=\frac{1}{2\sqrt{2}} \epsilon_{ijk}\frac{\langle j, k \rangle}{\sqrt{p_{123}^2}} \qquad ,\qquad i,j,k =1,2,3
\nonumber\\
 x^\pm_{i}&=\frac{\pm i}{2\sqrt{2}} \epsilon_{ijk}\frac{\langle j,k \rangle}{\sqrt{p_{123}^2}} \qquad,\qquad i,j,k =4,5,6
\end{align}
In order to determine the unknown functions $f^{\pm}(\l)$, we extract from the general expression (\ref{superamp4}) the purely scalar and the purely fermionic components and equal these expressions to our explicit results (\ref{tree}, \ref{scalar2}, \ref{fermion2})\footnote{In \cite{BLM} the $f^{\pm}(\l)$ functions were determined at tree level
using a different set of amplitudes, that is $(\phi_4 \bar \phi_4)^3$ and $(\psi_4 \bar \psi_4)^3$.}.

At tree level the purely fermionic six--point amplitude is vanishing, whereas the scalar one is a constant. Conversely, at one loop the latter is null, whilst the former is almost constant, up to discontinuities in the factor (\ref{fermion2}), which we have denoted by ${\cal C}(P)$.

Therefore, we obtain the following systems of equations for $f^+$ and $f^-$ at tree and one--loop level
\beq
\label{systems}
{\rm Tree:} \; \left\{ \begin{array}{l} A\, f^{+(0)} + B\, f^{-(0)} = 1 
 \\
A\, f^{+(0)} -  B\, f^{-(0)} = 0 \end{array}\right.
\qquad {\rm One-loop:} \;
\left\{ \begin{array}{l} A\, f^{+(1)} + B\, f^{-(1)} = 0 \\
A\, f^{+(1)} - B\, f^{-(1)} = -i \, {\cal C}(P)  \end{array}\right.
\eeq
where
\begin{align}
A &= -\frac{\left(\langle 1 \,|\, p_{56} \,|\, 4\rangle +i\, \langle 2\,3\rangle\,  \langle 5\,6\rangle\right) \left(\langle 3 \,|\, p_{45} \,|\, 6\rangle +i\, \langle 1\,2\rangle \, \langle 4\,5\rangle \right)  }{2 \sqrt{2}\, \sqrt{p^2_{123}}} \non\\
\non\\
B &= -\frac{\left(\langle 1 \,|\, p_{56} \,|\, 4\rangle -i\, \langle 2\,3\rangle \, \langle 5\,6\rangle\right) \left(\langle 3 \,|\, p_{45} \,|\, 6\rangle -i\, \langle 1\,2\rangle \, \langle 4\,5\rangle \right)  }{2 \sqrt{2}\, \sqrt{p^2_{123}}}
\end{align}

The solutions to the systems (\ref{systems}) are given in Appendix \ref{app:ffunctions}.

At tree level, plugging the result (\ref{soltree}) into the general form of the superamplitude as given in (\ref{general}) and re--expressing everything in terms of the $\eta^A$ variables, after some algebra we obtain
\bea
\mathcal{A}^{(0)}_6
&=& 
-\frac{\delta^3(P)\,\delta^6(Q)}{p_{123}^2} \left[\frac{ \left(\epsilon_{ijk}\, \langle j\,k \rangle\, \eta_i^I  + i\, \epsilon_{\bar{i}\bar{j}\bar{k}}\, \langle \bar{j}\, \bar{k} \rangle\, \eta_{\bar{i}}^I \right)^3}
{\left(\langle 1\,|\, p_{23} \,|\,4\rangle -i\, \langle 2\,3\rangle\,  \langle 5\,6\rangle \right) \left(\langle 3\,|\,p_{12}\,|\,6\rangle- i\, \langle 1\,2\rangle\,  \langle 4\,5\rangle \right)} \right. \nonumber
\\
&&\hspace{1cm}
\left. +\frac{ \left( \epsilon_{ijk}\, \langle j\,k \rangle\, \eta_i^I - i\, \epsilon_{\bar{i}\bar{j}\bar{k}}\, \langle \bar{j}\, \bar{k} \rangle\, \eta_{\bar{i}}^I \right)^3}
{\left(\langle 1\,|\,p_{23}\,|\,4\rangle +i\, \langle 2\,3\rangle \, \langle 5\,6\rangle \right) \left(\langle 3\,\,|p_{12}\,|\,6\rangle + i\, \langle 1\,2\rangle \, \langle 4\,5\rangle\right)}\right] \non \\
\label{modified}
\eea
As a check of this expression, one can easily see that our six--scalar and six--fermion amplitudes (\ref{tree}) are correctly reproduced, as well as the mixed amplitude $A(\bar{ \phi} \phi \bar{ \phi} \phi \bar{\psi} \psi )$ computed in \cite{GHKLL}.

At one loop level, the functions $f^{\pm}$ are given in eq. (\ref{solloop}). Plugging them in the general expression of the superamplitude and performing algebraic manipulations  similar to the tree level case,  the result reads
\bea
\mathcal{A}^{(1)}_6
&=& i\, 
{\cal C}(P)\,\, \frac{\delta^3(P)\,\delta^6(Q)}{p_{123}^2} \left[\frac{ \left(\epsilon_{ijk}\, \langle j\,k \rangle \,\eta_i^I  + i\, \epsilon_{\bar{i}\bar{j}\bar{k}}\, \langle \bar{j}\, \bar{k} \rangle\, \eta_{\bar{i}}^I \right)^3}
{\left(\langle 1\,|\, p_{23} \,|\,4\rangle -i\, \langle 2\,3\rangle \, \langle 5\,6\rangle \right) \left(\langle 3\,\,|p_{12}\,|\,6\rangle- i\, \langle 1\,2\rangle \, \langle 4\,5\rangle \right)} \right. \nonumber
\\
&&\hspace{1cm}
\left. -\frac{ \left( \epsilon_{ijk}\, \langle j\,k \rangle\, \eta_i^I - i\, \epsilon_{\bar{i}\bar{j}\bar{k}}\, \langle \bar{j}\, \bar{k} \rangle\, \eta_{\bar{i}}^I \right)^3}
{\left(\langle 1\,|\,p_{23}\,|\,4\rangle +i\, \langle 2\,3\rangle \, \langle 5\,6\rangle \right) \left(\langle 3\,|\,p_{12}\,|\,6\rangle + i\, \langle 1\,2\rangle \, \langle 4\,5\rangle\right)}\right] \non \\
\label{modified1L}
\eea
The analysis of this result and the discussion of its properties are postponed to Section 5.

\section{Generalization to $n$ points}

We now consider general  ${\rm N}^{(n/2 - 2)}$MHV amplitudes of the form (\ref{4n+2}, \ref{4n}) with
$n=4m+2$ or $n=4m$, respectively. Implementing a procedure similar to the one adopted for the six--point amplitude, we evaluate them up to one loop. 
 
 \subsection{Tree level}

At tree level, we depict the amplitudes as in Fig. \ref{fig:tree} where we have chosen the superfield on the middle line of the leftmost vertex to be conventionally $B_1$. We find convenient to assign momentum $p_0$ to that field and label the rest of momenta counterclockwise.  

\FIGURE{
  \centering
\includegraphics[width = 0.9 \textwidth]{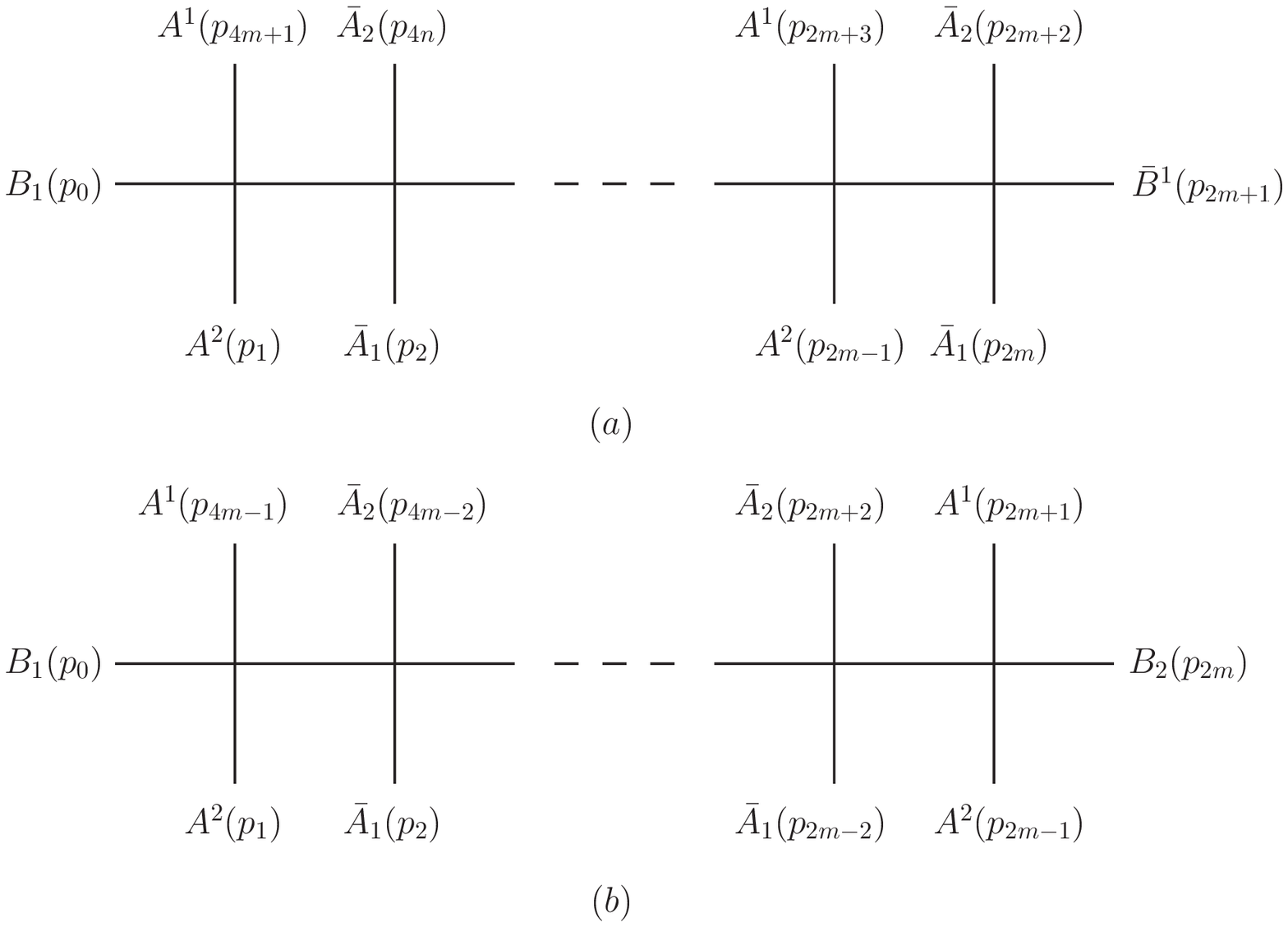}
\caption{Diagrams contributing to the tree level amplitudes with $(4m+2)$ and $4m$ external particles, respectively.}
\label{fig:tree}
}
The first diagram corresponds to the $(4m+2)$--point amplitude, whereas the second one to the $4m$--point amplitude.
 
In order to obtain effective action contributions from which we may read different component amplitudes,  
we temporarily work in Euclidean superspace and perform D--algebra on the two supergraphs.  

For the time being we forget about color indices and order the superfields in the most convenient way for making general formulae readable. At the end of the calculation, after going to components,  the fields will be reshuffled in order to take the planar order. This will give rise to possible signs from permutation of fermions.

As a result of performing D--algebra, the tree level effective action for $(4m+2)$ superfields looks like
\begin{align}
\label{effaction1}
& {\Gamma}^{tree}_{4m+2} \, \rightarrow \, i^{2m}\,
\int d^4\, \th\, A^1(p_{4m+1})\, B_1(p_0) \,A^2(p_1) \, \times \non\\&~~~~~~~~~~~~ \qquad\left[ \prod_{i=1}^{m-1}\, \bar A_1(p_{2i})\, \bar A_2(p_{4m+2-2i})\, D^2 \, A^2(p_{2i+1})\, A^1(p_{4m+1-2i}) \, \bar D^2\,  \right]\, \times \non\\&
~~~~~~~~~~~~ \qquad \quad \bar A_1(p_{2m})\, \bar B^1(p_{2m+1})\, \bar A_2(p_{2m+2})
\end{align}
whereas for  $4m$ superfields  
\begin{align}
\label{effaction2}
& {\Gamma}^{tree}_{4m} \, \rightarrow \, i^{2m-1}\,
\int d^4\, \th\, A^1(p_{4m-1})\, B_1(p_0) \,A^2(p_1) \, \times 
\non\\&
~~~~~~~~~~~\qquad  \left[ \prod_{i=1}^{m-2}\, \bar A_1(p_{2i})\, \bar A_2(p_{4m-2i})\, D^2 \, A^2(p_{2i+1})\, A^1(p_{4m-1-2i}) \, \bar D^2\,  \right] D^2\, \times \non\\&
~~~~~~~~~~~ \qquad  \qquad \bar A_1(p_{2m-2})\, \bar A_2(p_{2m+2})\,
A^2(p_{2m-1})\, B_2(p_{2m})\, A^1(p_{2m+1})\, 
\end{align}
where in both cases we have omitted the internal propagators which will be recovered when deriving the amplitudes.  In these formulae the $D^2$ and $\bar D^2$ operators have to be understood as acting on every field appearing on the right, and the factors inside the products are ordered from left to right according to increasing $i$ labels.  Eq. (\ref{effaction1}) is strictly valid for $m>1$; for $m=1$ the product inside square brackets has to be meant to be equal to 1. Similarly, eq. (\ref{effaction2}) makes sense for $ m>2$, whereas for $m=2$ the product has to be understood as equal to 1. The $i$ factors come from the internal vertices.  

From these expressions, component amplitudes can be obtained by performing the $d^4 \th$ integration. 
By a simple counting of derivatives, and taking into account the definition (\ref{components}) of the field components and the on--shell conditions (\ref{onshell}), we can infer preliminary information about their nature.  

We begin by discussing $(4m+2)$--point amplitudes. In eq. (\ref{effaction1}) there are $(2m+1)$ chiral and $(2m+1)$ antichiral superfields, while only  $2m$ $D$ and $2m$ $\bar D$ derivatives appear. The first obvious consequence is that purely fermionic amplitudes can never be generated. Instead, the maximally fermionic amplitude contains $4m$ fermions and 2 scalars. By suitably performing D--algebra, it is easy to see that the scalars always appear at the corners of the diagram, leading to nine possible such amplitudes. On the other hand, the fact that there is an equal number of chiral and antichiral derivatives, guarantees that it is always possible to perform D--algebra so as to obtain a purely scalar amplitude.  

For  $4m$--point amplitudes, the expression (\ref{effaction2}) contains $(2m+2)$ chiral superfields and $(2m-2)$ antichiral ones. In this case the number of $D$ and $\bar D$ derivatives, including those from the integration measure, are $2m$ and $(2m-2)$, respectively. Being them unequal, means that it is never possible to distribute spinorial derivatives in such a way to get a purely scalar amplitude that, therefore, vanishes. On the other hand, applying the spinorial derivatives to the greatest number of superfields compatibly with their (anti)chiral nature, will lead to amplitudes for $(4m-2)$ fermions and two scalars. The two scalars, coming from two chiral superfields, can only appear one on the left and one on the right extremal vertices, so leading to nine different configurations.

After these preliminary observations, we now extract from the effective actions (\ref{effaction1}, \ref{effaction2}) the explicit expressions for the simplest component amplitudes, that is the ones with the maximal number of scalars or fermions. 

We first focus on the $(4m+2)$--point case. As already discussed, non--vanishing purely scalar amplitudes can be obtained in this case by applying the same number of $D$ and $\bar{D}$ derivatives to (anti)chiral superfields and using the (anti)chirality conditions $\bar{D}_\a D_\b \Phi(p)  | =  p_{\a\b} \phi(p) $ and $D_\a \bar{D}_\b \bar{\Phi}(p)| = p_{\a\b} \bar{\phi}(p)$.  This produces square momentum factors at numerator which will cancel some of the propagators. Given the particular distribution of spinorial derivatives in (\ref{effaction1}), it is not difficult to prove by induction that the internal propagators which will be canceled are the ones corresponding to odd positions in Fig. $2(a)$. 

Therefore, going back to ${\cal N}=3$ notation for the scalar fields, we obtain   
\begin{align}
&\,\,\,\mathcal{A}^{(0)}_{4m+2} (A^1(p_{4m+1}), B_1(p_0), A^2(p_1), \bar{A}_1(p_2), \dots ,\bar{B}^1(p_{2m+1}), \cdots , \bar A_2 (p_{4m})) \, \equiv
\nonumber\\ 
\non \\
&
\mathcal{A}^{(0)}_{4m+2} (\phi^1(p_{4m+1}),\bar \phi_3(p_0), \phi^2(p_1) , \bar{\phi}_1(p_2) , \cdots 
\phi^3(p_{2m+1}), \cdots \bar \phi_2 (p_{4m}))  \non \\
&
\, \qquad = \, \prod_{i=1}^{m-1}\, \frac{1}{p^2_{4m+2-2i;1+4i}}
\end{align}
where every index is understood to be cyclic with period $(4m+2)$ and we have used the definition 
(\ref{nmomenta}) for the square of the sum of on--shell momenta.

Similarly, the almost completely fermionic amplitude may be easily obtained by applying the spinorial derivatives on the maximal number of superfields. Trading the spinorial fields with their polarization spinors, e.g. $D_\a A^i | \to \l_\a^i$, the contractions arising from the fermions associated to the internal vertices lead to the following factor  
\begin{equation}
\prod_{i=2}^{2m-1}\, \langle i , 4m+2-i \rangle
\end{equation}
Extra contractions will arise from fermions lying at the corner vertices, but these will be different according to the position we choose for the two scalars. 

If we focus on one particular amplitude, the one where the scalars are associated to the $B$ superfields, and go back to ${\cal N}=3$ notation, we obtain
\begin{align}
&\,\,\mathcal{A}^{(0)}_{4m+2} (\a^1(p_{4m+1}), B_1(p_0), \a^2(p_1) \dots \bar \a_1(p_{2m})\, \bar B^1(p_{2m+1})\, \bar \a_2(p_{2m+2}) \dots \bar \a_2 (p_{4m})) \equiv  
\non\\ 
\non \\
&\,\,
\mathcal{A}^{(0)}_{4m+2} (\psi_2(p_{4m+1}), \bar\phi_3(p_0), \psi_1(p_1) \dots \bar\psi^2(p_{2m})\, \phi^3(p_{2m+1})\, \bar \psi^1(p_{2m+2}) \dots \bar \psi^1 (p_{4m})) 
\non\\
& \,\,
\qquad = \, - \, i^{2m}\, 
\prod_{i=1}^{2m-1}\, \frac{1}{p^2_{4m+2-i;1+2i}}\, \prod_{i=1}^{2m}\, \langle i , 4m+2-i \rangle
\end{align}

As mentioned before, for $4m$ external particles the purely scalar amplitude of the type we are studying is always vanishing. For maximally fermionic amplitudes, the analysis goes through similarly to the $(4m+2)$ case.  Therefore, for instance, the almost completely fermionic amplitude with scalars associated to the $B$ superfields is
\begin{align}
&\,\, \mathcal{A}^{(0)}_{4m} (\a^1(p_{4m-1}), B_1(p_0), \a^2(p_1) \dots \a^2(p_{2m-1})\, B_2(p_{2m})\, \a^1(p_{2m+1}) \dots \bar \a_2 (p_{4m-2})) \equiv  
\non\\
\non \\ 
& - i\, \mathcal{A}^{(0)}_{4m} (\psi_2(p_{4m-1}), \bar\phi_3(p_0), \psi_1(p_1) \dots \psi_1(p_{2m-1})\, \bar\phi_4(p_{2m})\, \psi_2(p_{2m+1}) \dots \bar \psi_1 (p_{4m-2})) 
\non\\
&
\qquad = -i^{2m-1}\, 
\prod_{i=1}^{2m-2}\, \frac{1}{p^2_{4m-i;1+2i}}\, \prod_{i=1}^{2m-1}\, \langle i , 4m-i \rangle
\end{align}

\vskip 20pt
\subsection{One--loop}

Given the particular configurations of external  fields that we are considering, it is easy to realize that in the planar limit one--loop corrections to the superamplitudes (\ref{4n+2}, \ref{4n}) are simply given by diagrams similar to the ones in Fig. \ref{fig:6points}  for the six--point case, with the vector propagator connecting two adjacent matter lines in all possible ways.  

It is then sufficient to evaluate the contribution of generic blocks as the ones drawn in Fig. 
\ref{fig:blocks}, representing the insertion of the vector propagator at position $i$. 
\FIGURE{ 
    \centering
    \includegraphics[width=0.9\textwidth]{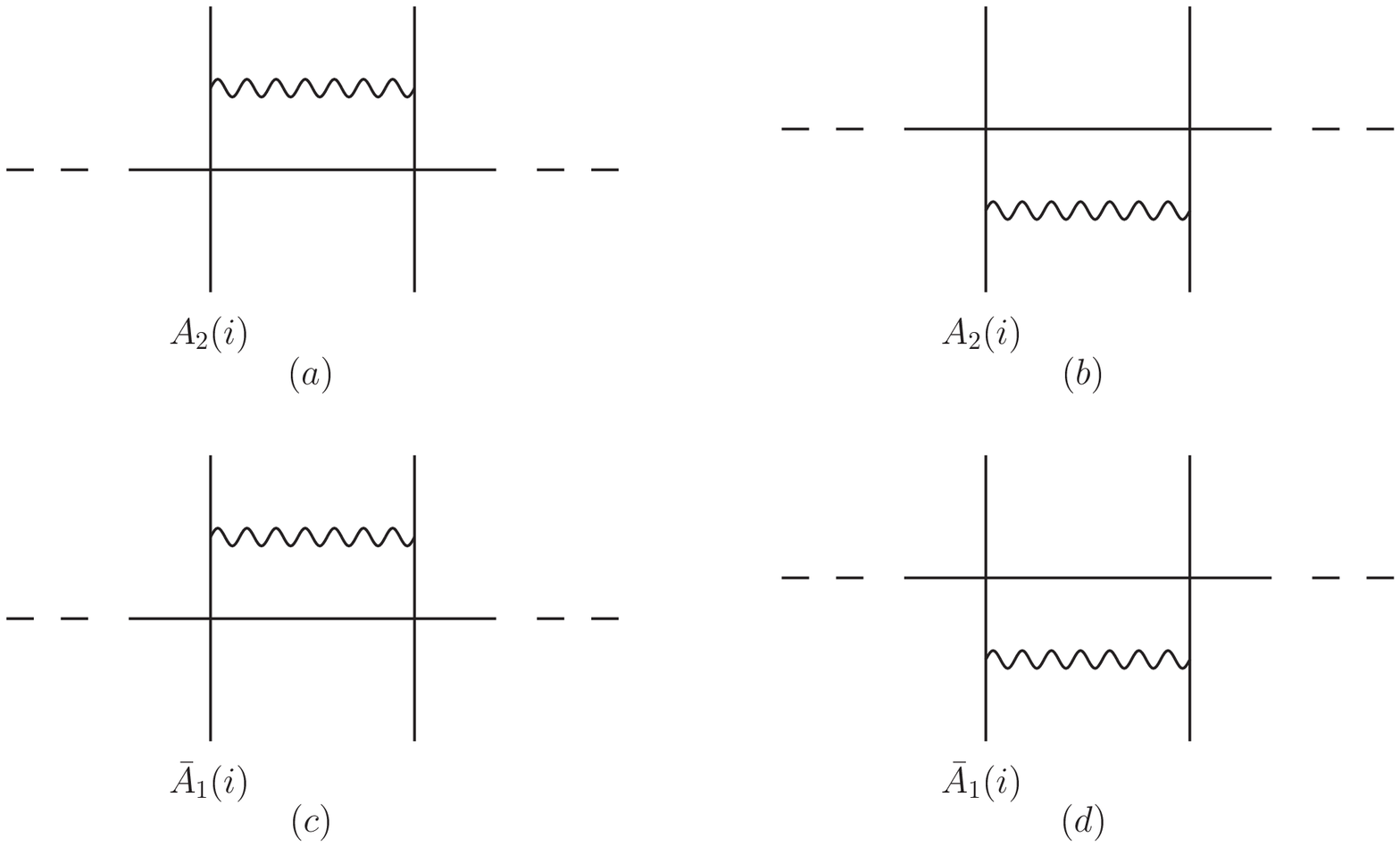}
    \caption{Blocks contributing at one--loop. In blocks $(a)$ and $(d)$ wavy lines correspond to $V$--vectors, whereas in blocks $(b)$ and $(c)$ they correspond to $\hat{V}$--vectors.}
    \label{fig:blocks}
}

We concentrate on a particular block, see Fig. \ref{fig:generalization}, where we have generically indicated (anti)chiral superfields with $\Phi$ ($\bar{\Phi}$) and introduced a label $n$ which can take values $(4m+2)$ or $4m$, according to the amplitude we are considering.  
\FIGURE{ 
    \centering
    \includegraphics[width=0.6\textwidth]{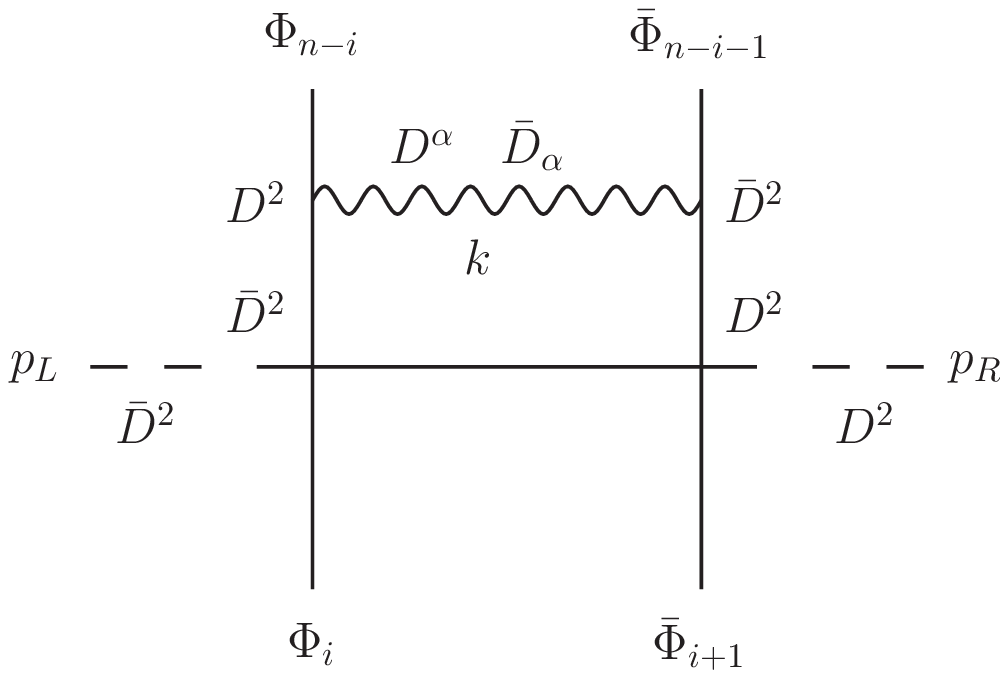}
    \caption{D--algebra for block ${\cal B}^{(a)}$.}
    \label{fig:generalization}
}
D--algebra leads to an expression similar to (\ref{intermediate}) for the six--point case, given by the sum of a triangle and a box momentum integrals. Forgetting for the moment the color indices and reordering the superfields according to convenience, we find 
\begin{align}
\label{intermediate2}
{\cal B}^{(a)}_i & = 
 i^{\frac{n}{2}-1} \, 4\p  \l\, (p_L+p_{n-i}+p_i)^2\,
\Delta(P)\, \int d^4\theta \, (\dots)_L\, \frac{d^3 k}{(2\pi)^3}\,\nonumber\\&  
\frac{\left[ k^2\, \,
D^{\a}\, \Phi_{n-i}\, \Db_{\a}\, \bar \Phi_{n-i-1}\, \Phi_{i}\, \bar \Phi_{i+1}
 + 2\, \varepsilon_{\mu\nu\rho}\, k^{\mu}\, p_{n-i}^{\nu}\, p_{n-i-1}^{\rho}\,
\Phi_{n-i}\, \bar \Phi_{n-i-1}\, \Phi_{i}\, \bar \Phi_{i+1}\right]}{k^2(k-p_{n-i})^2(k-p_L-p_i-p_{n-i})^2(k+p_{n-i-1})^2}
\, (\dots)_R\,   
\end{align}
where $\Delta(P)$ is the product of tree--level propagators entering the amplitude
\beq
\Delta(P) = \prod_{i=1}^{n/2-2}\, \frac{1}{p^2_{n-i;1+2i}}
\eeq
and $(\dots)_L$ and $(\dots)_R$ stand for the strings of fields and spinorial derivatives at the left and at the right of the block diagram. Their explicit expressions will depend on the position $i$ where the block is inserted, as well as on the kind of amplitude we are considering.    

Evaluating the momentum integral in (\ref{intermediate2}) and performing some non--trivial manipulation, the contribution from this block may be simplified to
\begin{align}
{\cal B}^{(a)}_i& =-\, i^{\tfrac{n}{2}-1}\, \frac{\p}{2}\, \l\,
 \Delta(P)\int d^4\theta\,  
\frac{
\Db^2\, D^2 \left[ (\dots)_L\, \Phi_i\, D^{\a}\, \Phi_{n-i} \right]\, \Db_{\a}\, \bar \Phi_{n-i-1}\, \bar \Phi_{i+1}
\, (\dots)_R}
{\sqrt{p_{L,i}^2}\,\sqrt{p_{n-i,n-i-1}^2}\,\sqrt{p_{i+1,R}^2}}
\non \\
\end{align}

Computing in a similar way the contributions from the other blocks in Fig. \ref{fig:blocks} we eventually find  
\begin{align}
{\cal B}^{(b)}_i& = i^{\tfrac{n}{2}-1}\, \frac{\p}{2}\,\hat\l\,
 \Delta(P)\int d^4\theta\,  
\frac{
\Db^2\, D^2 \left[ (\dots)_L\, \Phi_{n-i}\, D^{\a}\, \Phi_{i}\right]\, \Db_{\a}\, \bar \Phi_{i+1}\, \bar \Phi_{n-i-1}
\, (\dots)_R}
{\sqrt{p_{R,n-i-1}^2}\,\sqrt{p_{n-i,L}^2}\,\sqrt{p_{i,i+1}^2}}
\nonumber\\
{\cal B}^{(c)}_i& = i^{\tfrac{n}{2}-1}\, \frac{\p}{2}\,\hat\l\,
 \Delta(P)\int d^4\theta\,  
\frac{
D^2\, \Db^2 \left[ (\dots)_L\, \bar \Phi_i\, \bar D^{\a}\, \bar \Phi_{n-i} \right]\, D_{\a}\, \Phi_{n-i-1}\, \Phi_{i+1}
\, (\dots)_R}
{\sqrt{p_{R,n-i-1}^2}\,\sqrt{p_{n-i-1,n-i}^2}\,\sqrt{p_{i+1,R}^2}}
\nonumber\\
{\cal B}^{(d)}_i& = -i^{\tfrac{n}{2}-1}\, \frac{\p}{2}\,\l\,
 \Delta(P)\int d^4\theta\,  
\frac{
D^2 \,\Db^2 \left[ (\dots)_L\, \bar \Phi_{n-i}\, \bar D^{\a}\, \bar \Phi_{i} \right]\, D_{\a}\, \Phi_{i+1}\, \Phi_{n-i-1}
\, (\dots)_R}
{\sqrt{p_{L,n-i}^2}\,\sqrt{p_{i,i+1}^2}\,\sqrt{p_{n-i-1,R}^2}}
\end{align}

The full one--loop contribution is then given by a sum over these blocks, properly inserted in the corresponding tree level diagrams.
 
In the $(4m+2)$ case there are $m$ insertions of blocks $(a)$ and $(b)$, corresponding to odd indices and $(m-1)$ insertions of blocks $(c)$ and $(d)$, corresponding to even indices in the sum. Therefore, we can write
\begin{align}\label{eq:insertions1}
\Gamma^{1-loop}_{4m+2} \left( A^1(p_{4m+1}) B_1(p_0) \dots \bar A_2(p_{4m}) \right) = \sum_{i=1}^m\, \left( {\cal B}^{a}_{2i-1} + {\cal B}^{b}_{2i-1} \right)  +  \sum_{i=1}^{m-1}\, \left( {\cal B}^{c}_{2i} + {\cal B}^{d}_{2i} \right)
\non \\
\end{align}
The strings of fields on the left and on the right of the diagram are explicitly given by
\begin{align}
\label{dots1}
\left(\dots\right)_L & = \begin{cases}
\left[\prod_{j=1}^{(i-1)/2} \bar D^2\, \bar A_1(p_{i+1-2j}) \, \bar A_2(p_{4m+1-i+2j})\, D^2\, A_2(p_{i-2j})\, A_1(p_{4m+2-i+2j}) \right]\!\! 
B_1(p_0)& \\ & \hspace{-1.5cm} i~odd \\
\left[\prod_{j=1}^{(i-2)/2}\,
D^2\, A_2(p_{i+1-2j}) \, A_1(p_{4m+1-i+2j})\, \bar D^2\, \bar A_1(p_{i-2j})\, \bar A_2(p_{4m+2-i+2j}) 
\right]\, & \\
~~~~~~~~~~~~  D^2\, \left(A_1(p_{4m+1})\, B_1(p_0)\, A_2(p_{1})\right) & \hspace{-1.7cm} i~even
\end{cases} \non\\
\non \\
\left(\dots\right)_R & = \begin{cases}
\left[\prod_{j=1}^{m-(i+1)/2}\, D^2\, A_2(p_{i+2j}) \, A_1(p_{4m+2-i-2j})\, 
\bar D^2\, \bar A_1(p_{i+1+2j})\, \bar A_2(p_{4m+1-i-2j}) \right]\,& \\ ~~~~~~~~~~~~   \bar B_1(p_{2m+1}) & \hspace{-.95cm} i~odd \\
\left[\prod_{j=1}^{m-(i+2)/2}\,
\bar D^2\, \bar A_1(p_{i+2j}) \, \bar A_2(p_{4m+2-i-2j})\, 
D^2\, A_2(p_{i+1+2j})\, A_1(p_{4m+1-i-2j}) \right]\, & \\
~~~~~~~~~~~~  \bar D^2\, \left(\bar A_1(p_{2m})\, \bar B_1(p_{2m+1})\, \bar A_2(p_{2m+2})\right) &\hspace{-1.2cm}  i~even
\end{cases}
\end{align}
In the $4m$ case there are $(m-1)$ insertions of blocks $(a)$ and $(b)$, corresponding to odd indices and 
$(m-1)$ insertions of blocks $(c)$ and $(d)$, corresponding to even indices in the sum. We can write
\begin{align}\label{eq:insertions2}
\Gamma^{1-loop}_{4m} \left( A_1(p_{4m+1}) B_2(p_0) \dots \bar A_2(p_{4m}) \right) = \sum_{i=1}^{m-1}\, \left( {\cal B}^{a}_{2i-1} + {\cal B}^{b}_{2i-1} \right)  +  \sum_{i=1}^{m-1}\, \left( {\cal B}^{c}_{2i} + {\cal B}^{d}_{2i} \right)
\end{align}
The strings of fields on the left and on the right of the diagram are
\begin{align}
\label{dots2}
\left(\dots\right)_L & = \begin{cases}
\left[\prod_{j=1}^{(i-1)/2}\, \bar D^2\, \bar A_1(p_{i+1-2j}) \, \bar A_2(p_{4m-1-i+2j})\, D^2\, A_2(p_{i-2j})\, A_1(p_{4m-i+2j}) \right]\, 
B_1(p_0) & \\ & \hspace{-1.3cm} i~odd \vspace{1mm}\\
\left[\prod_{j=1}^{(i-2)/2}\, 
D^2\, A_2(p_{i+1-2j}) \, A_1(p_{4m-1-i+2j})\, \bar D^2\, \bar A_1(p_{i-2j})\, \bar A_2(p_{4m-i+2j}) \right]\, & \\
~~~~~~~~~~~~  D^2\, \left(A_1(p_{4m-1})\, B_1(p_0)\, A_2(p_{1})\right) & \hspace{-1.5cm} i~even 
\end{cases} \non\\ 
\non\\
\left(\dots\right)_R & = \begin{cases}
\left[\prod_{j=1}^{m-(i+3)/2}\, D^2\, A_2(p_{i+2j}) \, A_1(p_{4m-i-2j})\, 
\bar D^2\, \bar A_1(p_{i+1+2j})\, \bar A_2(p_{4m-1-i-2j}) \right]
& \\ ~~~~~~~~~~~~ 
\, D^2\, \left( A_2(p_{2m-1})\, B_2(p_{2m})\, A_1(p_{2m+1}) \right) & \hspace{-1.4cm} i~odd \\
\left[\prod_{j=1}^{m-(i+2)/2}\! 
\bar D^2 \bar A_1(p_{i+2j}) \bar A_2(p_{4m-i-2j}) 
D^2 A_2(p_{i+1+2j}) A_1(p_{4m-1-i-2j}) \right]
\!\! B_2(p_{2m}) &\\ & \hspace{-1.6cm} i~even 
\end{cases} 
\end{align}
In principle, the formulae above allow to extract all the component amplitudes within the particular class we are considering. Clearly, the procedure becomes more and more cumbersome as the number of particles grows, but this problem can be overcome by implementing the extraction of the components by a computer program.

Nevertheless, there exist some component amplitudes which are particularly simple, as they receive very few corrections. In fact, at tree level we concentrated on purely scalar and maximally fermionic amplitudes as the two cases where the number of possibilities to distribute spinorial derivatives on the superfields
gets minimized. 

At one loop, it can be easily inferred from D--algebra that in the class of amplitudes under investigation purely scalar amplitudes can never be generated.
Furthermore it is easy to realize that there are no purely fermionic amplitudes for $n > 6$. This is due to the fact that whenever a block is inserted, factors like $D^2 \left( A^1 B_2 A^2 \right)$ or $\bar D^2 \left( \bar A_1 \bar B^2 \bar A_2 \right)$ get produced, which can never lead to three fermions.
On the other hand, whenever the block is inserted at the corners of the diagram this is partially avoided. In fact,  in these cases it is possible to place three fermions on the left corner or, similarly, at the right one. This is the technical reason why at six points a purely fermionic amplitude arises.

Exploiting this pattern, we can restrict to a subclass of amplitudes with $4m$ ($4m-2$) fermions, out of the external $4m+2$ ($4m$) superfields, where for instance we require all fields at the left corner to be fermions. These are particularly simple cases because the amplitudes receive quantum corrections by two diagrams only, the ones with upper and lower blocks at the left corner, independently of the number of external legs.   

Explicitly, the two contributions from $(a)$ and $(b)$ blocks divided by their tree level counterpart read
\begin{align}
& -\frac{\p}{2}\,\l\, \int d^4\theta\,
\frac{\Db^2\, D^2 \left[ D^{\a}\, A_1(p_{n-1})\, B_1(p_0)\, A_2(p_1) \right]\, \bar A_1(p_2)
\, (\dots)_R\,  \Db_{\a}\, \bar A_2(p_{n-2})}
{\sqrt{p_{01}^2}\,\sqrt{p_{n-1,n-2}^2}\,\sqrt{p_{n-2;4}^2}}
\non\\ &
+ \frac{\p}{2}\,\hat\l\, \int d^4\theta\,
\frac{
\Db^2\, D^2 \left[ A_1(p_{n-1})\, B_1(p_0)\, D^{\a}\, A_2(p_1)\right]\, \Db_{\a}\, \bar A_1(p_2)
\, (\dots)_R \, \bar A_2(p_{n-2})}
{\sqrt{p_{n-1;4}^2}\,\sqrt{p_{0, n-1}^2}\,\sqrt{p_{12}^2}}
\end{align}
Performing the $\theta$--integration while requiring the fields at the left corner to be fermions, leads necessarily to apply the $D^2$ operator from the measure on the first term in square brackets. In fact, this gives rise to $D^2 \bar{D}^2 D^2 [\cdots] = -p^2_{n-1;3} \,D^2 [\cdots]$, and distributing the remaining $D^2$ on the three superfields in the brackets in the only possible way compatible with the equations of motion (\ref{onshell}), we obtain three fermions.

The remaining $\bar D^2$ from the measure must act on the rest of the string of superfields. There, the ellipses $(\cdots)_R$ depend on the number of total particles, but they are always given by a string beginning with a $D^2$ operator acting on the fields and derivatives on the right of the block (see eqs. (\ref{dots1}, \ref{dots2})).
Depending on how we distribute the derivatives on the superfields, different component amplitudes will be generated.  

As an example, we select the amplitude where all the superfields give rise to fermions, except for $A^2(p_3)$ and the $\bar B^1$ (or $B_2$) superfield at the right corner, which give rise to the two scalars. This fixes D--algebra uniquely and we obtain the general expression valid for $n \geq 8$
\begin{align}\label{eq:npt}
& {\cal A}^{(1)}\left(\a^1 (p_{n-1})\,  \b_1 (p_{0})\, \a^2 (p_{1})\, \bar \a_1 (p_{2})\, A^2 (p_{3})\, \bar \a_1 (p_{4}) \dots B(p_{(n-2)/2}) \dots \bar \a_2(p_{n-2}) \right)
= \non\\ & =
- \, \frac{\p}{2}\, p^2_{n-1;3}\, \prod_{i=3}^{(n-2)/2}\, \langle i,n-i \rangle \, \times
\\ &
\left[ \l\,
\frac{\langle 0 \, 1 \rangle}{\sqrt{\langle 0 \,1 \rangle^2}} \, \frac{\langle n-2 , n-1 \rangle}{\sqrt{\langle n-2 ,n-1 \rangle^2}} \,
\frac{\langle 2 \, 3 \rangle}{\sqrt{p_{n-2;4}^2}}
- \hat\l\,
\frac{\langle 1 \,2 \rangle}{\sqrt{\langle 1 \,2 \rangle^2}} \, \frac{\langle n-1 , 0 \rangle}{\sqrt{\langle n-1 , 0 \rangle^2}} \, \frac{\langle 3 , n-2 \rangle}{\sqrt{p_{n-1;4}^2}} \right]
\non
\end{align}
where the $B$ scalar  stands either for $\bar B^1$ or $B_2$. 
 
This result shares the peculiar features of the six--point amplitudes, namely the presence of discontinuities due to the ratios of invariants producing sign functions.

\section{Discussion and conclusions}

We now discuss  important properties of the one--loop amplitudes we have found.  

First of all, we have determined one--loop non--vanishing amplitudes for any number $n$ of external particles with $n \geq 6$. The result is given in terms of scalar triangle integrals, the only one--loop topology that exhibits dual conformal covariance. 

These amplitudes, in contrast with the four--point ones, are never MHV and, in analogy with the four dimensional case, are not expected to be dual to bosonic light--like polygon Wilson loops. For this reason, there is no contradiction between our findings and the one--loop vanishing of $n$--polygon Wilson loops in any three dimensional Chern--Simons theory, with or without matter. 

In four dimensional ${\cal N}=4$ SYM theory, ${\rm N}^k {\rm MHV}$ amplitudes have been shown to be dual to the $\th^{4k}$--component of a superWilson loop constructed in terms of ${\cal N}=4$ superconnections integrated over light--like closed paths in ${\cal N}=4$ superspace \cite{Mason:2010yk, CaronHuot:2010ek, Belitsky:2011zm}. 
It would be interesting to find an analogous construction for a superWilson loop in ABJ(M) theories whose components should be dual to the amplitudes we have computed. The construction might be complicated by the fact that in three dimensions we do not have an on--shell ${\cal N}=6$ superspace description.   

We now discuss Yangian invariance of  our amplitudes at one loop. For simplicity, we concentrate on the full six--point superamplitude of Section 3. 

For ${\cal N}=4$ SYM in four dimensions, the generators of the superconformal algebra under which superamplitudes are invariant, are identified as the level zero generators of a Yangian algebra. Level one generators are constructed as bilocal composites of level zero generators and give rise to dual superconformal symmetry of the scattering amplitudes. The closure of the level zero and one generators forms the Yangian algebra \cite{DNW1,DNW2}. 

For three dimensional theories the analogous construction has been worked out in \cite{BLM} where the explicit expression for level zero and one generators has been given. Successively, in analogy with the four dimensional case, it has been shown that level one generators of Yangian algebra are equivalent to dual superconformal generators when acting on on--shell amplitudes \cite{HL}. 

As already mentioned, at any loop order superconformal invariance restricts the form of the six--point superamplitude to be \cite{BLM}
\beq
{\cal A}^{(l)}_6 = \d^3(P)\, \d^6(Q)\, \left[ f^{+(l)}(\l) \, \d^3(\a) + f^{-(l)}(\l) \, \d^3(\b) \right]
\eeq
This has been used in Section 3 for determining the exact expression of the superamplitude at tree and one--loop order, starting from the knowledge of two components. 
 
At tree level, requiring the six--point superamplitude in question to be annihilated by the superconformal and dual superconformal generators translates into two constraints on the $f^{+(0)}(\lambda_i)$ and $f^{-(0)}(\lambda_i)$ functions, which come in the form of differential equations in the spinor variables $\lambda_i^{\alpha}$.
In \cite{BLM} it was shown that these are satisfied separately by both functions for generic choices of the external momenta, implying Yangian invariance of the superamplitude at this order. 

We then study what happens at one loop.
A first important observation is that from our explicit evaluation it is easy to realize that the one--loop functions $f^{\pm(1)}$ are proportional to the tree--level ones $f^{\pm(0)}$, according to 
\beq
f^{+(1)}=-i\,\mathcal{C}(P)\,f^{+(0)}\quad\mbox{and}\quad f^{-(1)}=i\,\mathcal{C}(P)\,f^{-(0)}
\eeq
where the proportionality factor $\mathcal{C}(P)$ has been given in eq. (\ref{fermion2}).   

The interesting feature of $\mathcal{C}(P)$ is that it is given as a sum of products of $\langle k l\rangle/\sqrt{\langle k l\rangle^2}$ factors. Using the prescription (\ref{root}) for the determination of square roots of momentum invariants, these factors give rise to sign functions that
evaluate to $\pm 1$, depending on the particular kinematic configurations of the external particles. It follows that the $\mathcal{C}(P)$ factor is always constant and can take four different values $\pm \frac{\pi}{2} \,(\lambda+\hat\lambda)$ or $\pm \frac{\pi}{2} \,(\lambda-\hat\lambda)$, which reduce to $\pm\pi\lambda$ or strictly $0$ in the ABJM case. 

The $\mathcal{C}(P)$ factor inherits discontinuities from the sign functions. Since these are discontinuous in points where their argument goes to zero, it follows that the regions of discontinuities of $\mathcal{C}(P)$ are represented by sets of points where two particle momenta become collinear. 

Away from collinear configurations the $f^{\pm(1)}$ functions are separately proportional to the tree--level ones, up to a constant factor. Therefore, the one--loop superamplitude inherits Yangian invariance from its tree level counterpart. This happens locally, in all regions of constancy for the $\mathcal{C}(P)$ factor.

Careful analysis has to be devoted to collinear kinematic regions where $\mathcal{C}(P)$ becomes discontinuous. In fact, close to these regions the differential equations required by Yangian invariance get spoiled by the appearance of an anomaly. 

To understand how the anomaly arises, we first observe that given the definition (\ref{root}) for a generic
$\langle k l\rangle/\sqrt{\langle k l\rangle^2}$ ratio, the direct application of the derivative with respect to one of the two spinors involved in the contraction gives
\begin{equation}\label{eq:spinder}
\frac{\partial}{\partial\lambda_k^{\a}}\left(\frac{\langle k l\rangle}{\sqrt{\langle k l\rangle^2}}\right)=
2a \, \lambda_{l\a}\,\delta\left(a \, \langle k l \rangle\right)
\end{equation}
where $a = 1$ for $E_l E_k<0$, whereas $a = -i$ for $E_l E_k>0$.  

The appearance of the $\delta$--function has drastic consequences when applying the superconformal generators $\mathfrak{S}^A_{\a} = \eta^A \frac{\pa}{\pa \l^\a}$ to the  six--point superamplitude. In fact, the variation of the six--point superamplitude under these generators reads \cite{BLM}
\begin{equation}
\mathfrak{S}^A_{\a}\, {\cal A}_6
=\delta^3(P)\, \delta^6(Q)\left(
	 \left( \sum_{k=1}^6\, x^+_k\, \frac{\partial\, f^{+} }{\partial\, \lambda^{\a}_k} \right) \beta^A\delta^3(\alpha)
	+ \left\{(\alpha,+)\leftrightarrow(\beta,-)\right\} \right)
\label{eq:anomaly}
\end{equation}
where $x^{\pm}$ have been defined in (\ref{pm}).

Focusing on the one--loop superamplitude, the variation of the ${\cal C}(P)$ factor inside $f^{\pm(1)}$ gives\begin{equation}
\mathfrak{S}^A_{\a}\, {\cal A}_6^{(1)} 
= \delta^3(P)\, \delta^6(Q)\left(
	 \left( -i\,f^{+(0)}\, \sum_{k=1}^6\, x^+_k\, \frac{\partial\, {\cal C}(P) }{\partial\, \lambda^{\a}_k} \right) \beta^A\delta^3(\alpha)
	- \left\{(\alpha,+)\leftrightarrow(\beta,-)\right\} \right)
\label{eq:anomaly2}
\end{equation}
To evaluate this expression we restrict for instance to a configuration where scattered particles have alternating energy signs, so to keep all spinor contractions of adjacent momenta real. All other energy configurations can be adjusted adding $i$ factors as mentioned above.

Recalling the explicit expression (\ref{fermion2}) for the ${\cal C}(P)$ factor and using the identity (\ref{eq:spinder}), the term $\sum_{k=1}^6\, x^+_k\, \frac{\partial\, {\cal C}(P) }{\partial\, \lambda^a_k}$ turns out to be non--trivial and given by (all indices are understood to be cyclic $mod$ 6)
\begin{align}
\sum_{k=1}^6\, x^+_k\, \frac{\partial\, {\cal C}(P) }{\partial\, \lambda^{\a}_k} &=
\pi\, \l \, 
\sum_{i = 1,3,5}
x^+_{[ i}\, \lambda_{i+1]\, \a}\, \delta (\langle i,i+1\rangle )\, 
 \frac{\langle i+2,i+3\rangle}{\sqrt{\langle i+2,i+3\rangle^2}}\, \frac{\langle i+4,i+5\rangle}{\sqrt{\langle i+4,i+5\rangle^2}}
 \non\\&
+ \pi\, \hat \l \, 
\sum_{i = 2,4,6}
x^+_{[ i}\, \lambda_{i+1]\, \a}\, \delta (\langle i,i+1\rangle )\, 
 \frac{\langle i+2,i+3\rangle}{\sqrt{\langle i+2,i+3\rangle^2}}\, \frac{\langle i+4,i+5\rangle}{\sqrt{\langle i+4,i+5\rangle^2}}
\end{align}
Therefore the superconformal generators $\mathfrak{S}^A_{\a}$ act non--trivially on the superamplitude whenever we are close to configurations that correspond to $\delta$--function supports, that is, at collinear limits. This signals the presence of an anomaly in the variation of the one--loop superamplitude that strictly resembles the holomorphic anomaly occurring at tree level in four dimensions \cite{Cachazo:2004by}--\cite{Bargheer:2011mm}. 

Similarly, the level one generator of the Yangian algebra $\mathfrak{P}^{(1)}$ constructed in \cite{BLM} can be shown not to annihilate the 
$\mathcal{C}(P)$ function either. Therefore, dual superconformal symmetry is also anomalous at one loop. 

For scattering processes involving more than six particles we have found only few component amplitudes, while an expression for the complete superamplitude is still lacking. However, already  at component level, 
the particular $n$--point amplitudes we have computed (see eq. (\ref{eq:npt})) exhibit the same kind of discontinuities as the ones of the six--point case. Therefore, a pattern similar to the one described above is expected to emerge and will lead to the appearance of anomalies.  

In four dimensions, the tree--level holomorphic anomaly arising in $n$--point amplitudes can be written as an operator acting on a $(n-1)$--point amplitude. At the level of generating functional of all the amplitudes the exact invariance can then be recovered by deforming the classical superconformal generators via the addition of this extra operator \cite{Bargheer:2009qu}--\cite{Bargheer:2011mm}.
It would be interesting to investigate whether a similar pattern might be implemented in three dimensions in order to cancel the one--loop anomaly that we have found.  

In four dimensions, delta--function anomalies at tree level can be efficiently used for computing amplitudes at one loop. In fact, for MHV and NMHV amplitudes they have been exploited at the unitarity cuts in order to determine the coefficients of one--loop box integrals \cite{Cachazo:2004dr, Britto:2004nj, Bidder:2004tx}. Along the same line of reasoning, the one--loop anomaly we have found could be exploited for computing two--loop amplitudes via generalized unitarity cuts.

\section*{Acknowledgements}

We greatly acknowledge discussions with Till Bargheer, Niklas Beisert, Florian Loebbert and Tristan McLoughlin.

This work has been supported in part by INFN and MIUR--PRIN contract 2009--KHZKRX, and
by the research grants MICINN-09-FPA2009- 07122 and MEC-DGI-CSD2007-00042.

\vfill
\newpage
\appendix
\section{Conventions and results in three dimensions}

The results for the amplitudes are given in Minkowski metric $g_{\mu\nu}={\rm diag}\{-1,1,1\}$.

Fourier transform to momentum space is defined as
\beq
f(x)  = \int \frac{d^3p}{(2\pi)^3} \, e^{-ipx} \, \tilde{f}(p) 
\eeq
from which $i \pa_{\a\b} \to p_{\a\b}$.

On--shell solutions of the fermionic equations of motion are expressed in terms of $SL(2,\mathbb{R})$ commuting spinors $\l_\a$. The same quantities allow to write on--shell momenta as 
\beq 
\label{momentum}
p_{\alpha\beta} \equiv p_{\mu}(\gamma^{\mu})_{\alpha\beta} =\lambda_{\alpha}\lambda_{\beta}
\eeq
where the set of $2 \times 2$ gamma matrices are chosen to satisfy 
\beq
\left(\g^{\mu}\right)^{\a\g}\, \left(\g^{\nu}\right)_{\g\b} = g^{\m\n}\, \d^{\a}_{~\b} - i\, \e^{\m\n\rho}\, \left(\g_{\rho}\right)^{\a}_{~\b}
\eeq
An explicit set of matrices is $( \gamma^{\mu} )^{\a\b}= \{ \s^0, \s^3, \s^1 \}$. 
  
Spinorial indices are raised and lowered as  
\begin{eqnarray}
  \l^\alpha=C^{\alpha\beta}\l_\beta  \qquad \l_\alpha=\l^\beta C_{\beta\alpha}
\end{eqnarray}
where the $C$ matrix is
\begin{eqnarray}
  C^{\alpha\beta} = \left(\begin{array}{cc} 0 & i \\ -i & 0 \end{array}\right) \qquad
  C_{\alpha\beta} = \left(\begin{array}{cc} 0 & -i \\ i & 0 \end{array}\right)
\end{eqnarray}
We define contractions as
\beq
\langle i \, j\rangle=-\langle j \, i\rangle \equiv \lambda^{\alpha}_i\lambda_{\alpha j}=C^{\alpha\beta}
\lambda_{\beta i}\lambda_{\alpha j}
\eeq
For any couple of on--shell momenta we write
\beq
p_{ij}^2 \equiv (p_i+p_j)^2 = 2 \, p_i \cdot p_j =  p_i^{\a\b} \, (p_j)_{\a\b} = \langle i \, j\rangle^2
\eeq
Analogously, for three of them we use
\beq
p_{ijk}^2 \equiv (p_i+p_j+p_k)^2 = 2 \, p_i \cdot p_j + 2 \, p_i \cdot p_k + 2 \, p_j \cdot p_k =  \langle i \, j\rangle^2 + \langle i \, k\rangle^2 + \langle j \, k\rangle^2 
\eeq
More generally, we define 
\beq
\label{nmomenta}
p^2_{i;j}=\left( \sum\limits_{k=0}^{j-1}p_{i+k} \right)^2
\eeq
Inverting eq. (\ref{momentum}), the energy of the particle is given by
\beq
E \equiv p^0 = \frac12\, (\g^0)^{\a\b} p_{\a\b} = \frac12\left(\l_1^{\,2} + \l_2^{\,2}\right) 
\eeq
It follows that real spinors describe positive energy solutions of the Dirac equation, that is particles traveling forward in time, whereas purely imaginary spinors represent negative energy solutions, that is particles traveling backwards in time. 

As a consequence, given particles $k$ and $l$, the nature of their polarization spinors, or equivalently the signs of their energies,  determines the sign of the two--particle invariant $(p_k + p_l)^2$ as follows:
If $E_k E_l < 0$ then one of the spinors is real and the other one is purely imaginary. Therefore, $\langle k \, l \rangle$ is real and its square $\langle k \, l\rangle^2 = (p_k + p_l)^2$ positive.
On the other hand, if $E_k E_l > 0$, then the two spinors have the same nature, their contraction is purely imaginary and its square is negative. 

We define square roots of two--particle invariants via the $i \e$ prescription. Depending on the sign of the energies we have 
\begin{align}
\label{root}
&\frac{\langle k l\rangle}{\sqrt{\langle k l\rangle^2+i\e}}=\text{Sign}\left[\langle k l \rangle\right]\quad \quad ~\mbox{for}\quad E_k E_l<0\nonumber\\
&\frac{\langle k l\rangle}{\sqrt{\langle k l\rangle^2+i\e}}=\text{Sign}\left[-i\,\langle k l \rangle\right]\quad \mbox{for}\quad E_k E_l>0 
\end{align}
where Sign is the sign function. Note that the '$i$' factor inside the argument of the second sign function compensates the fact that in this case the $\langle k l \rangle$ contraction is purely imaginary. The whole argument of the Sign function is thus real and well defined in both cases.

\vskip 25pt

For one--loop calculations we have used the following massive triangle 
\begin{equation}
\label{integral1}
{\cal T}(p_{im}, p_{jl}) = \int\frac{d^3k}{(2\pi)^3}\frac{1}{(k-p_i)^2\,(k-p_i - p_j - p_l)^2\,(k+p_m)^2}=\frac{1}{8\sqrt{p_{im}^2} \sqrt{p_{jl}^2}\sqrt{p_{r}^2}}
\end{equation}
and a tensorial box integral which can be re--expressed in terms of the scalar triangle
\bea
\label{integral2}
{\cal Q}(p_i, p_{jl}, p_m) &=& \int\frac{d^3k}{(2\pi)^3}\frac{\e_{\mu\nu\rho}\, k^{\mu}\,p_{i}^{\nu}\,p_{m}^{\rho}}{k^2\,(k-p_i)^2\,(k-p_i -p_j -p_l)^2\,(k+p_m)^2}
\\
\non \\
&=& \frac{\e_{\mu\nu\rho} \, (p_j+p_l)^{\mu} \,p_{i}^{\nu}\,p_{m}^{\rho}}{8\,(p_i+p_j+p_l)^2 \sqrt{p_{im}^2} \sqrt{p_{jl}^2}\sqrt{p_{r}^2}} = 
 \frac{\e_{\mu\nu\rho}\,p_{jl}^{\mu} \, p_{i}^{\nu}\, p_{m}^{\rho}}{p_{ijl}^2} \, {\cal T}(p_{im}, p_{jl})
 \non 
\eea
where $p_{r} = - p_{im} -p_{jl}$. The last equality can be also proved at the level of Feynman--parametrized integrals.

\section{The ABJ(M) theory in ${\cal N}=2$ notation}

A realization of  ${\cal N}=6$ supersymmetric ABJ(M) models can be given in terms of
${\cal N}=2$ three dimensional superspace \cite{Klebanov}.  For $U(M) \times U(N)$
gauge group, the physical field content is organized into two vector multiplets $(V,\hat{V})$
in the adjoint representation of the first and the second group respectively, 
coupled to chiral multiplets $A^i$ and $B_i$ carrying a fundamental index $i=1,2$ of a global $SU(2)_A \times SU(2)_B$ and in the bifundamental and antibifundamental representations of the gauge group, respectively.

To derive effective action contributions from which we extract amplitudes, we work in euclidean superspace $(x^{\a\b}, \th^\a, \bar{\th}^\b)$, $\a, \b = 1,2$, with the effective action defined as $e^{\G} = \int e^S$. The ${\cal N}=6$ supersymmetric action reads
\begin{equation}
 {\cal S} = {\cal S}_{\mathrm{CS}} + {\cal
    S}_{\mathrm{mat}}
  \label{eqn:action}
\end{equation}
with
\begin{align}
  \label{action}
& {\cal S}_{\mathrm{CS}}
=  \frac{K}{4\pi} \, \int d^3x\,d^4\theta \int_0^1 dt\: \Big\{  \Tr \Big[
V\, \Db^\a \left( e^{-t V}\, D_\a\, e^{t V} \right) \Big]
-\Tr \Big[ \hat{V}\, \Db^\a \left( e^{-t \hat{V}} D_\a
e^{t\, \hat{V}}\, \right) \Big]   \Big\} \nonumber \\
& {\cal S}_{\mathrm{mat}} = \int d^3x\,d^4\theta\: \Tr \left( \bar{A}_i\,
e^V\, A^i\, e^{- \hat{V}} + \bar{B}^i\, e^{\hat V}\, B_i\,
e^{-V} \right)\nonumber \\
&  +\frac{2\pi i}{K}\int d^3x\, d^2\theta\, \epsilon_{ik}\,\epsilon^{jl}\,
\mathrm{Tr}\,\left(A^i\, B_j\, A^k\, B_l\right)+\frac{2\pi i}{K}\int d^3x\, d^2\bar\theta\,
\epsilon^{ik}\,\epsilon_{jl}\, \mathrm{Tr}\,\left(\bar A_i\, \bar B^j\, \bar A_k\, \bar B^l\right)
\end{align}
Here $K$  is an integer, as required by gauge invariance of the effective action.
In the perturbative regime we take $\l \equiv \frac{M}{K} \ll 1$ and $\hat{\l} \equiv \frac{N}{K} \ll 1$.

Superspace covariant derivatives are defined as
\begin{equation}
  D_\a = \pa_\a + \frac{i}{2}\,  \thb^\b\,  \pa_{\a\b}
  \qquad , \qquad \Db_\a = \bar\pa_\a
  + \frac{i}{2}\,  \th^\b\,  \pa_{\a\b}
\end{equation}
and satisfy $\{D_\a ,\,  \Db_\b\} = i\,  \pa_{\a\b}$. 

We require the external particles to be on--shell, that is to satisfy the free equations of motion
\beq
\label{onshell}
D^2 A^i = D^2 B_i = 0 \qquad , \qquad \bar{D}^2 \bar{A}_i = \bar{D}^2 \bar{B}^i = 0
\eeq
The quantization of the theory can be easily carried on in superspace
after performing gauge fixing (for details, see for instance
\cite{BPS}). In momentum space and using Landau gauge, this
leads to gauge propagators
\begin{eqnarray}
  \langle V^a_{\, b}(1) \, V^c_{\, d}(2) \rangle
  =   \frac{4\pi}{K} \, \frac{1}{p^2} \,  \, \delta^a_d \, \d^c_b \times \Db^\a D_\a \, \delta^4(\th_1-\th_2) \nonumber \\
  \langle \hat V^{\bar{a}}_{\bar{b}} (1) \, \hat V^{\bar{c}}_{\bar{d}}(2) \rangle =-
  \frac{4\pi}{K} \,  \frac{1}{p^2} \, \,   \delta^{\bar{a}}_{\bar{d}} \, \d^{\bar{c}}_{\bar{b}} \times \Db^\a D_\a  \, \delta^4(\th_1-\th_2)
  \label{gaugeprop}
\end{eqnarray}
whereas the matter propagators are
\begin{eqnarray}
  &&\langle \bar A^{\bar{a}}_{\ a}(1) \, A^b_{\ \bar{b}}(2) \rangle
  = \frac{1}{p^2} \,\, \delta^{\bar{a}}_{\ \bar{b}} \, \delta^{\ b}_{a} \times  D^2 \bar{D}^2 \, \delta^4(\th_1 - \th_2)
 \nonumber \\
  &&  \langle \bar B^a_{\ \bar{a}}(1) \, B^{\bar{b}}_{\ b}(2) \rangle =
   \frac{1}{p^2} \,\, \delta^a_{\ b} \, \delta^{\ \bar{b}}_{\bar{a}} \times D^2 \bar{D}^2 \, \delta^4(\th_1 - \th_2)
\label{scalarprop}
\end{eqnarray}
where $a,b$ and $\bar{a}, \bar{b}$ are indices of the fundamental representation of the first and the second gauge groups, respectively.
The vertices employed in our one--loop calculation can be easily read from
the action (\ref{action}) and they are given by
\begin{eqnarray}
  \label{vertices}
 && \int d^3x\,d^4\theta\: \left[ \Tr ( \bar{A}_i V A^i) -  \Tr ( B_i V \bar{B}^i )
    + \Tr (  \bar{B}^i {\hat V} B_i ) -  \Tr ( A^i {\hat{V}}   \bar{A}_i ) \right]
    \non \\
  && \qquad + \frac{4\pi i}{K} \int d^3x\,d^2\theta\:
    \, \Big[ \Tr (A^1 B_1 A^2 B_2) -  \Tr (A^1 B_2 A^2 B_1)\Big] ~+~ {\rm h.c.}
\end{eqnarray}

The field components are defined as
\bea
\label{components}
&& A^i | = A^i  \quad \quad , \quad B_i | = B_i  \quad , \quad \bar{A}_i | = \bar{A}_i  \quad \quad , \quad \bar{B}^i | = \bar{B}^i
\non \\
&& D A^i | = \a^i  \quad , \quad D B_i | = \b_i  \quad , \quad \bar{D} \bar{A}_i | = \bar{\a}_i \quad , \quad
\bar{D} \bar{B}^i | = \bar{\b}^i
\eea

\section{Six--point amplitude: The $f^{\pm}$ functions}\label{app:ffunctions}

In this appendix we list explicitly the $f^{\pm}$ functions entering the computation of the six--point superamplitude in Section \ref{sec:sixpoint}
\bea
\label{soltree}
f^{+(0)} &=& -
\frac{\sqrt{2} \sqrt{p_{123}^2}}{\left(\langle 1\,|\, p_{23} \,|\,4\rangle -i\, \langle 2\,3\rangle \, \langle 5\,6\rangle \right) \left(\langle 3\,|\,p_{12}\,|\,6\rangle- i\, \langle 1\,2\rangle \, \langle 4\,5\rangle \right)} \non\\
f^{-(0)} &=& -
\frac{\sqrt{2} \sqrt{p_{123}^2}}{\left(\langle 1\,|\,p_{23}\,|\,4\rangle +i\, \langle 2\,3\rangle \, \langle 5\,6\rangle \right) \left(\langle 3\,|\,p_{12}\,|\,6\rangle + i\, \langle 1\,2\rangle \, \langle 4\,5\rangle\right)}
\eea
Similarly, the $f^{\pm}$ functions at one loop read
\beq
\label{solloop}
f^{+(1)} = -i\, 
{\cal C}(P)\, f^{+(0)} \quad,\quad
f^{-(1)} = i\, 
{\cal C}(P)\, f^{-(0)}
\eeq
where ${\cal C}(P)$ has been defined in eq. (\ref{fermion2}). 


\vfill
\newpage


\begin{thebibliography}{99}

\bibitem{ABJM}
  O.~Aharony, O.~Bergman, D.~L.~Jafferis and J.~Maldacena,
  {\em ``N=6 superconformal Chern-Simons-matter theories, M2-branes and their
  gravity duals''}, 
  JHEP {\bf 0810} (2008) 091,
  [arXiv:0806.1218 [hep-th]].

\bibitem{ABJ}
 O.~Aharony, O.~Bergman and D.~L.~Jafferis,
  {\em ``Fractional M2-branes''}, 
  JHEP {\bf 0811} (2008) 043,
  [arXiv:0807.4924 [hep-th]].

\bibitem{Arutyunov:2008if}
  G.~Arutyunov and S.~Frolov,
   {\em``Superstrings on AdS(4) x CP**3 as a Coset Sigma-model''},
  JHEP {\bf 0809} (2008) 129
  [arXiv:0806.4940 [hep-th]].
  
\bibitem{Stefanski:2008ik}
  B.~Stefanski, jr,
  {\em``Green-Schwarz action for Type IIA strings on AdS(4) x CP**3''},
  Nucl.\ Phys.\ B {\bf 808} (2009) 80
  [arXiv:0806.4948 [hep-th]].
  
\bibitem{Sorokin:2010wn}
  D.~Sorokin and L.~Wulff,
  {\em``Evidence for the classical integrability of the complete $AdS_4 x CP^3$ superstring''},
JHEP {\bf 1011} (2010) 143 [arXiv:1009.3498 [hep-th]].
  
\bibitem{Kalousios:2009ey}
  C.~Kalousios, C.~Vergu and A.~Volovich,
  {\em``Factorized Tree-level Scattering in AdS(4) x CP**3''},
  JHEP {\bf 0909} (2009) 049
  [arXiv:0905.4702 [hep-th]].

\bibitem{MZ}
   J.~A.~Minahan, K.~Zarembo,
 {\em``The Bethe ansatz for superconformal Chern-Simons''},
  JHEP {\bf 0809 } (2008)  040.
  [arXiv:0806.3951 [hep-th]]; \\
  J.~A.~Minahan, W.~Schulgin, K.~Zarembo,
   {\em``Two loop integrability for Chern-Simons theories with N=6 supersymmetry''},
  JHEP {\bf 0903 } (2009)  057.
  [arXiv:0901.1142 [hep-th]].

\bibitem{GV}
  N.~Gromov, P.~Vieira,
   {\em``The all loop AdS4/CFT3 Bethe ansatz''},
  JHEP {\bf 0901 } (2009)  016.
  [arXiv:0807.0777 [hep-th]].

\bibitem{GV2}
  N.~Gromov and P.~Vieira,
   {\em``The AdS(4) / CFT(3) algebraic curve''},
  JHEP {\bf 0902} (2009) 040
  [arXiv:0807.0437 [hep-th]].

\bibitem{AN}
  C.~Ahn and R.~I.~Nepomechie,
   {\em``N=6 super Chern-Simons theory S-matrix and all-loop Bethe ansatz equations''},
  JHEP {\bf 0809} (2008) 010
  [arXiv:0807.1924 [hep-th]].

\bibitem{Astolfi:2011bg}
  D.~Astolfi, G.~Grignani, E.~Ser-Giacomi and A.~V.~Zayakin,
  {\em ``Strings in $AdS_4 x CP^3$: finite size spectrum vs. Bethe Ansatz''},
  JHEP {\bf 1204} (2012) 005
  [arXiv:1111.6628 [hep-th]].


\bibitem{Nishioka:2008gz}
  T.~Nishioka, T.~Takayanagi,
  {\em  ``On Type IIA Penrose Limit and N=6 Chern-Simons Theories''},
  JHEP {\bf 0808 } (2008)  001.
  [arXiv:0806.3391 [hep-th]].

\bibitem{Gaiotto:2008cg}
  D.~Gaiotto, S.~Giombi, X.~Yin,
  {\em  ``Spin Chains in N=6 Superconformal Chern-Simons-Matter Theory''},
  JHEP {\bf 0904 } (2009)  066.
  [arXiv:0806.4589 [hep-th]].

\bibitem{Grignani}
  G.~Grignani, T.~Harmark and M.~Orselli,
  {\em  ``The SU(2) x SU(2) sector in the string dual of N=6 superconformal
  Chern-Simons theory''},
  Nucl.\ Phys.\  B {\bf 810} (2009) 115
  [arXiv:0806.4959 [hep-th]].

\bibitem{MRT}
  T.~McLoughlin, R.~Roiban, A.~A.~Tseytlin,
  {\em  ``Quantum spinning strings in AdS(4) x CP**3: Testing the Bethe Ansatz proposal''},
  JHEP {\bf 0811 } (2008)  069
  [arXiv:0809.4038 [hep-th]].

\bibitem{GromovM}
  N.~Gromov, V.~Mikhaylov,
  {\em  ``Comment on the Scaling Function in AdS(4) x CP**3''},
  JHEP {\bf 0904 } (2009)  083
  [arXiv:0807.4897 [hep-th]].

\bibitem{Kris}
  C.~Krishnan,
  {\em  ``AdS(4)/CFT(3) at One Loop''},
  JHEP {\bf 0809 } (2008)  092
  [arXiv:0807.4561 [hep-th]].

\bibitem{McR}
  T.~McLoughlin, R.~Roiban,
  {\em  ``Spinning strings at one-loop in AdS(4) x P**3''},
  JHEP {\bf 0812 } (2008)  101
  [arXiv:0807.3965 [hep-th]].

\bibitem{AAB}
  L.~F.~Alday, G.~Arutyunov, D.~Bykov,
  {\em  ``Semiclassical Quantization of Spinning Strings in AdS(4) x CP**3''},
  JHEP {\bf 0811 } (2008)  089
  [arXiv:0807.4400 [hep-th]].
 
\bibitem{Mika}
  V.~Mikhaylov,
  {\em  ``On the Fermionic Frequencies of Circular Strings''},
  J.\ Phys.\ A {\bf A43 } (2010)  335401
  [arXiv:1002.1831 [hep-th]].

\bibitem{Abbott}
  M.~C.~Abbott, I.~Aniceto, D.~Bombardelli,
  {\em  ``Quantum Strings and the $AdS_4/CFT_3$ Interpolating Function''},
  JHEP {\bf 1012 } (2010)  040
  [arXiv:1006.2174 [hep-th]].

\bibitem{Astolfi}
  D.~Astolfi, V.~G.~M.~Puletti, G.~Grignani, T.~Harmark and M.~Orselli,
  {\em  ``Finite-size corrections for quantum strings on $AdS_4 x CP^3$''},
  JHEP {\bf 1105} (2011) 128
  [arXiv:1101.0004 [hep-th]].

\bibitem{MOS1}
  J.~A.~Minahan, O.~Ohlsson Sax, C.~Sieg,
  {\em``Anomalous dimensions at four loops in N=6 superconformal Chern-Simons theories''},
  Nucl.\ Phys.\  {\bf B846 } (2011)  542-606
  [arXiv:0912.3460 [hep-th]].

\bibitem{MOS2}
  J.~A.~Minahan, O.~Ohlsson Sax, C.~Sieg,
  {\em  ``Magnon dispersion to four loops in the ABJM and ABJ models''},
  J.\ Phys.\ A {\bf A43}, 275402 (2010)
  [arXiv:0908.2463 [hep-th]].

\bibitem{LMMSSST}
  M.~Leoni, A.~Mauri, J.~A.~Minahan, O.~O.~Sax, A.~Santambrogio, C.~Sieg, G.~Tartaglino-Mazzucchelli,
  {\em ``Superspace calculation of the four-loop spectrum in N=6 supersymmetric Chern-Simons theories''},
  JHEP {\bf 1012 } (2010)  074
  [arXiv:1010.1756 [hep-th]].

\bibitem{Drummond:2009fd}
  J.~M.~Drummond, J.~M.~Henn and J.~Plefka,
   {\em``Yangian symmetry of scattering amplitudes in N=4 super Yang-Mills theory''}
  JHEP {\bf 0905} (2009) 046
  [arXiv:0902.2987 [hep-th]].
    
\bibitem{Drummond:2007aua}
  G.~P.~Korchemsky, J.~M.~Drummond and E.~Sokatchev,
   {\em``Conformal properties of four-gluon planar amplitudes and Wilson loops''},
  Nucl.\ Phys.\  B {\bf 795} (2008) 385
  [arXiv:0707.0243 [hep-th]].

\bibitem{Brandhuber:2007yx}
  A.~Brandhuber, P.~Heslop and G.~Travaglini,
   {\em``MHV Amplitudes in N=4 Super Yang-Mills and Wilson Loops''},
  Nucl.\ Phys.\  B {\bf 794} (2008) 231
  [arXiv:0707.1153 [hep-th]].

\bibitem{Drummond:2007cf}
  J.~M.~Drummond, J.~Henn, G.~P.~Korchemsky and E.~Sokatchev,
   {\em``On planar gluon amplitudes/Wilson loops duality''},
  Nucl.\ Phys.\  B {\bf 795} (2008) 52
  [arXiv:0709.2368 [hep-th]].

\bibitem{AEKMS}
  L.~F.~Alday, B.~Eden, G.~P.~Korchemsky, J.~Maldacena and E.~Sokatchev,
   {\em``From correlation functions to Wilson loops''},
  JHEP {\bf 1109} (2011) 123
  [arXiv:1007.3243 [hep-th]].

\bibitem{EKS}
  B.~Eden, G.~P.~Korchemsky and E.~Sokatchev,
   {\em``From correlation functions to scattering amplitudes''},
  JHEP {\bf 1112} (2011) 002
  [arXiv:1007.3246 [hep-th]].
  

\bibitem{BLM}
  T.~Bargheer, F.~Loebbert, C.~Meneghelli,
   {\em``Symmetries of Tree-level Scattering Amplitudes in N=6 Superconformal Chern-Simons Theory''},
  Phys.\ Rev.\  {\bf D82 } (2010)  045016
  [arXiv:1003.6120 [hep-th]].

\bibitem{HL2}
  Y.~-t.~Huang and A.~E.~Lipstein,
   {\em``Amplitudes of 3D and 6D Maximal Superconformal Theories in Supertwistor Space''},
  JHEP {\bf 1010} (2010) 007
  [arXiv:1004.4735 [hep-th]].

\bibitem{HL}
  Y.~-t.~Huang, A.~E.~Lipstein,
   {\em``Dual Superconformal Symmetry of N=6 Chern-Simons Theory''},
  JHEP {\bf 1011 } (2010)  076
  [arXiv:1008.0041 [hep-th]].

\bibitem{Drummond:2008vq}
  J.~M.~Drummond, J.~Henn, G.~P.~Korchemsky and E.~Sokatchev,
   {\em``Dual superconformal symmetry of scattering amplitudes in N=4 super-Yang-Mills theory''},
  Nucl.\ Phys.\  B {\bf 828} (2010) 317
  [arXiv:0807.1095 [hep-th]].

\bibitem{GHKLL}
  D.~Gang, Y.~-t.~Huang, E.~Koh, S.~Lee and A.~E.~Lipstein,
   {\em``Tree-level Recursion Relation and Dual Superconformal Symmetry of the ABJM Theory''},
  JHEP {\bf 1103} (2011) 116
  [arXiv:1012.5032 [hep-th]].

\bibitem{BCFW}
  R.~Britto, F.~Cachazo, B.~Feng and E.~Witten,
   {\em``Direct proof of tree-level recursion relation in Yang-Mills theory''},
  Phys.\ Rev.\ Lett.\  {\bf 94} (2005) 181602
  [hep-th/0501052].

\bibitem{Lee}
  S.~Lee,
   {\em``Yangian Invariant Scattering Amplitudes in Supersymmetric Chern-Simons Theory''},
  Phys.\ Rev.\ Lett.\  {\bf 105} (2010) 151603
  [arXiv:1007.4772 [hep-th]].

\bibitem{ABM}
  A.~Agarwal, N.~Beisert, T.~McLoughlin,
   {\em``Scattering in Mass-Deformed N>=4 Chern-Simons Models''},
  JHEP {\bf 0906 } (2009)  045
  [arXiv:0812.3367 [hep-th]].

\bibitem{CH}
  W.~-M.~Chen and Y.~-t.~Huang,
   {\em``Dualities for Loop Amplitudes of N=6 Chern-Simons Matter Theory''},
  JHEP {\bf 1111} (2011) 057
  [arXiv:1107.2710 [hep-th]].

\bibitem{BLMPS1}
  M.~S.~Bianchi, M.~Leoni, A.~Mauri, S.~Penati and A.~Santambrogio,
   {\em``Scattering Amplitudes/Wilson Loop Duality In ABJM Theory''},
  JHEP {\bf 1201} (2012) 056
  [arXiv:1107.3139 [hep-th]].

\bibitem{BLMPS2}
  M.~S.~Bianchi, M.~Leoni, A.~Mauri, S.~Penati and A.~Santambrogio,
   {\em ``Scattering in ABJ theories''},
  JHEP {\bf 1112} (2011) 073
  [arXiv:1110.0738 [hep-th]].

\bibitem{HPW}
  J.~M.~Henn, J.~Plefka, K.~Wiegandt,
   {\em ``Light-like polygonal Wilson loops in 3d Chern-Simons and ABJM theory''},
  JHEP {\bf 1008 } (2010)  032
  [arXiv:1004.0226 [hep-th]].


\bibitem{BM}
  N.~Berkovits, J.~Maldacena,
   {\em Fermionic T-Duality, Dual Superconformal Symmetry, and the Amplitude/Wilson Loop Connection''},
  JHEP {\bf 0809 } (2008)  062
  [arXiv:0807.3196 [hep-th]].

\bibitem{ADO}
  I.~Adam, A.~Dekel, Y.~Oz,
   {\em On Integrable Backgrounds Self-dual under Fermionic T-duality''},
  JHEP {\bf 0904 } (2009)  120
  [arXiv:0902.3805 [hep-th]].

\bibitem{Grassi:2009yj}
  P.~A.~Grassi, D.~Sorokin, L.~Wulff,
   {\em ``Simplifying superstring and D-brane actions in $AdS_4 x CP^3$ superbackground''},
  JHEP {\bf 0908 } (2009)  060
  [arXiv:0903.5407 [hep-th]].

\bibitem{Adam:2010hh}
  I.~Adam, A.~Dekel, Y.~Oz,
   {\em ``On the fermionic T-duality of the $AdS_4 x CP^3$ sigma-model''},
  JHEP {\bf 1010 } (2010)  110
  [arXiv:1008.0649 [hep-th]].

\bibitem{Bakhmatov}
  I.~Bakhmatov,
   {\em ``On $AdS_4 x CP^3$ T-duality''},
  Nucl.\ Phys.\  {\bf B847 } (2011)  38-53
  [arXiv:1011.0985 [hep-th]].

\bibitem{DO}
  A.~Dekel and Y.~Oz,
   {\em ``Self-Duality of Green-Schwarz Sigma-Models''},
  JHEP {\bf 1103} (2011) 117
  [arXiv:1101.0400 [hep-th]].

\bibitem{Bakhmatov:2011aa}
  I.~Bakhmatov, E.~O.~Colgain and H.~Yavartanoo,
   {\em ``Fermionic T-duality in the pp-wave limit''},
  JHEP {\bf 1110} (2011) 085
  [arXiv:1109.1052 [hep-th]].

\bibitem{Colgain:2012ca}
  E.~OColgain,
   {\em ``Self-duality of the D1-D5 near-horizon''},
  arXiv:1202.3416 [hep-th].

\bibitem{AM}
  L.~F.~Alday, J.~M.~Maldacena,
   {\em ``Gluon scattering amplitudes at strong coupling''},
  JHEP {\bf 0706 } (2007)  064
  [arXiv:0705.0303 [hep-th]];
 {\em ``Comments on gluon scattering amplitudes via AdS/CFT''},
  JHEP {\bf 0711 } (2007)  068
  [arXiv:0710.1060 [hep-th]].

\bibitem{BLP}
  M.~S.~Bianchi, M.~Leoni and S.~Penati,
   {\em ``An All Order Identity between ABJM and N=4 SYM Four-Point Amplitudes''},
  arXiv:1112.3649 [hep-th].

\bibitem{BLMPRS}
  M.~S.~Bianchi, M.~Leoni, A.~Mauri, S.~Penati, C.~A.~Ratti and A.~Santambrogio,
   {\em``From Correlators to Wilson Loops in Chern-Simons Matter Theories''},
  JHEP {\bf 1106} (2011) 118
  [arXiv:1103.3675 [hep-th]].
  
\bibitem{BBLM}
  T.~Bargheer, N.~Beisert, F.~Loebbert and T.~McLoughlin,
  {\em ``Conformal Anomaly for Amplitudes in N=6 Superconformal Chern-Simons Theory''},
  arXiv:1204.4406 [hep-th].
  
\bibitem{Klebanov}
  M.~Benna, I.~Klebanov, T.~Klose and M.~Smedback,
   {\em ``Superconformal Chern-Simons Theories and $AdS_4/CFT_3$ Correspondence''},
  JHEP {\bf 0809}, 072 (2008)
  [arXiv:0806.1519 [hep-th]].
 
\bibitem{Mason:2010yk}
  L.~J.~Mason and D.~Skinner,
   {\em ``The Complete Planar S-matrix of N=4 SYM as a Wilson Loop in Twistor Space''},
  JHEP {\bf 1012} (2010) 018
  [arXiv:1009.2225 [hep-th]].

\bibitem{CaronHuot:2010ek}
  S.~Caron-Huot,
   {\em ``Notes on the scattering amplitude / Wilson loop duality''},
  JHEP {\bf 1107} (2011) 058
  [arXiv:1010.1167 [hep-th]].

\bibitem{Belitsky:2011zm}
  A.~V.~Belitsky, G.~P.~Korchemsky and E.~Sokatchev,
   {\em ``Are scattering amplitudes dual to super Wilson loops?''},
  Nucl.\ Phys.\ B {\bf 855} (2012) 333
  [arXiv:1103.3008 [hep-th]].

\bibitem{DNW1}
  L.~Dolan, C.~R.~Nappi and E.~Witten,
   {\em ``A Relation between approaches to integrability in superconformal
Yang-Mills theory''},
  JHEP {\bf 0310} (2003) 017
  [hep-th/0308089].
  
\bibitem{DNW2}
  L.~Dolan, C.~R.~Nappi and E.~Witten,
   {\em ``Yangian symmetry in D = 4 superconformal Yang-Mills theory''},
  hep-th/0401243.

\bibitem{Cachazo:2004by}
  F.~Cachazo, P.~Svrcek and E.~Witten,
   {\em ``Gauge theory amplitudes in twistor space and holomorphic anomaly''},
  JHEP {\bf 0410} (2004) 077
  [hep-th/0409245].
  
\bibitem{Bargheer:2009qu} 
  T.~Bargheer, N.~Beisert, W.~Galleas, F.~Loebbert and T.~McLoughlin,
   {\em ``Exacting N=4 Superconformal Symmetry''},
  JHEP {\bf 0911}, 056 (2009)
  [arXiv:0905.3738 [hep-th]].
  
\bibitem{Sever:2009aa}
  A.~Sever and P.~Vieira,
   {\em ``Symmetries of the N=4 SYM S-matrix''},
  arXiv:0908.2437 [hep-th].
  
\bibitem{Beisert:2010gn}
  N.~Beisert, J.~Henn, T.~McLoughlin and J.~Plefka,
   {\em ``One-Loop Superconformal and Yangian Symmetries of Scattering Amplitudes in N=4 Super Yang-Mills''},
  JHEP {\bf 1004} (2010) 085
  [arXiv:1002.1733 [hep-th]].
  
\bibitem{Bargheer:2011mm}
  T.~Bargheer, N.~Beisert and F.~Loebbert,
   {\em ``Exact Superconformal and Yangian Symmetry of Scattering Amplitudes''},
  J.\ Phys.\ A A {\bf 44} (2011) 454012
  [arXiv:1104.0700 [hep-th]].
   
\bibitem{Cachazo:2004dr}
  F.~Cachazo,
   {\em ``Holomorphic anomaly of unitarity cuts and one-loop gauge theory amplitudes''},
  hep-th/0410077.
  
\bibitem{Britto:2004nj}
  R.~Britto, F.~Cachazo and B.~Feng,
   {\em ``Computing one-loop amplitudes from the holomorphic anomaly of unitarity cuts''},
  Phys.\ Rev.\ D {\bf 71} (2005) 025012
  [hep-th/0410179].
  
\bibitem{Bidder:2004tx}
  S.~J.~Bidder, N.~E.~J.~Bjerrum-Bohr, L.~J.~Dixon and D.~C.~Dunbar,
   {\em ``N=1 supersymmetric one-loop amplitudes and the holomorphic anomaly of unitarity cuts''},
  Phys.\ Lett.\ B {\bf 606} (2005) 189
  [hep-th/0410296].

\bibitem{BPS}
  M.~S.~Bianchi, S.~Penati and M.~Siani,
   {\em ``Infrared stability of ABJ-like theories''},
  JHEP {\bf 1001} (2010) 080
  [arXiv:0910.5200 [hep-th]];
   {\em ``Infrared Stability of N=2 Chern-Simons Matter Theories''},
  JHEP {\bf 1005} (2010) 106
  [arXiv:0912.4282 [hep-th]].
  
\end{thebibliography}
\end{document}